\documentclass[9pt,twocolumn,twoside]{opticajnl}

\usepackage{hyperref}
\usepackage{physics}
\usepackage{lineno}
\usepackage{graphicx}
\usepackage[percent]{overpic}
\usepackage{subcaption}
\usepackage{placeins}
\usepackage{multirow}
\usepackage{rotating}
\usepackage{tablefootnote}
\usepackage{mathtools}
\usepackage{colortbl}
\usepackage{multirow}
\usepackage{array}
\usepackage{orcidlink}

\DeclareRobustCommand{\svdots}{
  \vbox{%
    \baselineskip=0.33333\normalbaselineskip
    \lineskiplimit=0pt
    \hbox{.}\hbox{.}\hbox{.}%
    \kern-0.2\baselineskip
  }%
}

\graphicspath{{./}{figures/}}

\renewcommand\thesubsection{\thesection.\Alph{subsection}}

\journal{opticajournal} 

\title{In-Band Scattering and Absorption of Infrared Blocking Foam Filters for Millimeter-wave Cameras}

\author[1,2,*]{Alex Thomas\,\orcidlink{0000-0001-9528-8147}}
\author[3]{Bugao Zou\,\orcidlink{0000-0002-5032-8587}}
\author[4]{Shreya Sutariya}
\author[5]{Yuhan Wang\,\orcidlink{0000-0002-8710-0914}}
\author[6, 7]{Gabriele Coppi\,\orcidlink{0000-0002-6362-6524}}
\author[8]{Samuel Day-Weiss\,\orcidlink{0009-0003-5814-2087}}
\author[9]{Nicholas Galitzki\,}
\author[10,1]{Kathleen Harrington\,\orcidlink{0000-0003-1248-9563}}
\author[2]{Erin Healy\,\orcidlink{0000-0002-3757-4898}}
\author[4,2]{Claire Lessler}
\author[4]{Aashrita Mangu\,\orcidlink{0009-0000-1028-3524}}
\author[1,2,4,11,12]{Jeffrey McMahon}
\author[3,13]{Michael D. Niemack\,\orcidlink{0000-0001-7125-3580}}
\author[14]{Edward J. Wollack\,\orcidlink{0000-0002-7567-4451}}

\affil[1]{Department of Astronomy and Astrophysics, University of Chicago, Chicago, IL 60637, USA}
\affil[2]{Kavli Institute for Cosmological Physics, University of Chicago, Chicago, IL 60637, USA}
\affil[3]{Department of Applied and Engineering Physics, Cornell University, Ithaca, NY, 14853, USA}
\affil[4]{Department of Physics, University of Chicago, Chicago, IL 60637, USA}
\affil[5]{Department of Physics, Cornell University, Ithaca, NY, 14853, USA}
\affil[6]{Dipartimento di Fisica - Università degli Studi Milano Bicocca, Italy}
\affil[7]{Istituto Nazionale di Fisica Nucleare, INFN, Sezione Milano-Bicocca, Italy}
\affil[8]{Joseph Henry Laboratories of Physics, Jadwin Hall, Princeton University, Princeton, NJ 08544, USA}
\affil[9]{Department of Physics, University of Texas at Austin, Austin, TX 78712, USA}
\affil[10]{High Energy Physics Division, Argonne National Laboratory, Lemont, IL 60439, USA}
\affil[11]{Enrico Fermi Institute, University of Chicago, 5640 S Ellis Ave, Chicago, IL, 60637, USA}
\affil[12]{Fermi National Accelerator Laboratory, Batavia, IL 60510, USA}
\affil[13]{Department of Astronomy, Cornell University, Ithaca, NY, 14853, USA}
\affil[14]{NASA Goddard Space Flight Center, 8800 Greenbelt Road, Greenbelt, MD 20771, USA}

\affil[*]{alexgt@uchicago.edu}

\begin{abstract}
Expanded closed-cell polymer foams are widely used as thermal infrared (IR) blocking filters in millimeter-wave cameras, particularly for Cosmic Microwave Background observations. Precise knowledge of their millimeter-wave properties is essential for optimizing sensitivity. We present broadband (150~GHz – 2~THz) transmittance spectroscopy of Styroace-II and several Zotefoam filters, fitting their spectra with a radiative transfer model incorporating dielectric absorption and Rayleigh, Mie, and higher-order scattering. For a typical 5~cm thick filter stack at 280~GHz, Styroace-II exhibits $\mathbf{{\sim}10\%}$ scattering with absorption estimated as $\mathbf{{\lesssim}5\%}$ by effective-medium theory, while Zotefoam HD30 offers superior performance at $\mathbf{{\sim}3\%}$ scattering and absorption likewise bounded to $\mathbf{{{\lesssim}0.3\%}}$. Each model component is constrained at the $\mathbf{{\sim}0.1\%}$ transmittance level for millimeter wavelengths. We observe batch-to-batch scattering variability of up to 2 percentage points in foams with multiple tested batches. 
Less commonly used Zotefoam formulations (LD15 and LD24) can further reduce in-band scattering to $\mathbf{{<}1\%}$ while maintaining negligible in-band absorption and likely comparable IR blocking due to shared polyethylene absorption features and similar cell sizes. 
Based on this work, a filter constructed from the best measured LD24 batch has replaced the Styroace-II filter in a Simons Observatory 220/280 GHz Small Aperture Telescope.
\end{abstract}
\setboolean{displaycopyright}{false} 

\begin{document}

\maketitle

\section{Introduction}

\label{sec:intro}

Superconducting millimeter-wave detectors operate in the cryogenic, low-noise regime where marginal increases in optical loading can degrade sensitivity and limit the science return.   Operating these detectors requires filters to block infrared (IR) radiation while transmitting the desired millimeter-wave frequencies.  
Optimizing the performance of modern millimeter surveys requires detailed characterization of the radiative transport mechanisms in all optical elements, in particular the IR blocking filters responsible for balancing optical throughput with minimal optical and thermal loading.

We focus on measurements of the cosmic microwave background (CMB), where ambitious sensitivity goals drive optical requirements \cite{SO_2020, Sobrin_2022, snowmass2021}.
To maintain the low-noise cryogenic operation, CMB experiments mitigate the thermal IR loading introduced by the warm environment and atmosphere by employing IR blockers.
One widely-adopted solution is expanded closed-cell polymer foams, which attenuate IR loading via a combination of scattering and absorption processes that may produce optical loading.
While these foams have been validated by successful deployments across generations of CMB experiments, a predictive physical understanding of their scattering and absorption has remained absent. Here we present high-fidelity broadband transmittance spectra and a radiative transfer model that connects foam microstructure to quantitative in-band scattering and absorption predictions. While this work offers a generalized exploration of the mm-wave properties of these foams, we use and make constraints for Simons Observatory (SO), a CMB experiment in the Atacama Desert.


The materials characterized in this work are Styroace-II (referred to throughout as Styrofoam) used in POLARBEAR \cite{POLARBEAR} and SO Small Aperture Telescopes (SATs) \cite{SAT_paper, Day-Weiss_2026}; and Zotefoam HD30, used in BICEP3/Keck \cite{BICEP_windows}, SPT3G \cite{Sobrin_2022}, and SO Large Aperture Telescope (LAT) \cite{SO_2020}. The foams considered are dielectric mixtures comprised of cells: low density voids (typically gaseous or vacuum) surrounded by polymer shells that provide structural integrity. We also study less commonly used Zotefoams of various densities. 

Because of their low effective refractive index, these foams possess minimal reflections and can be stacked together to create IR blocking filters. These filters are deployed after the window along the optical chain. Some implementations use spacers between layers while others do not. We approximate the stack as multi-layer insulators where the outer layer is in radiative equilibrium with the external environment and the inner-most layer with the next stages of the cryostat, an alumina filter at 40 K for Simons Observatory. Each layer thermalizes with the foam layers on either side of it. With tens of layers, the radiative thermal equilibrium attenuates thermal loading by more than an order of magnitude.


We discuss infrared blocking foams in \S\ref{sec:IRBFs}, present precision measurements of their transmission spectra in \S\ref{sec:measurement}, and fit a radiative transfer model that incorporates absorption from dielectric loss, Rayleigh scattering, Mie scattering, and secondary scattering coupling the two scattering pathways in \S\ref{sec:model}. We estimate microphysical properties along with optical and thermal loading in \S\ref{sec:ep}, discuss the implications for mm-wave cameras in \S\ref{sec:discussion}; and summarize our general conclusions in \S\ref{sec:conclusion}. Appendix~\hyperref[appendix:A]{A} provides the underlying scattering and radiative transfer theory; Appendix~\hyperref[appendix:B]{B} derives the scattering efficiencies for closed-cell foams; Appendix~\hyperref[appendix:C]{C} includes scattering kernel simulations and off-axis measurements; Appendix~\hyperref[appendix:D]{D} outlines the mathematics for higher order corrections, and Appendix~\hyperref[appendix:E]{F} lists atmospheric waterline contamination masked in our data processing.

\section{Infrared Blocking Foams}
\label{sec:IRBFs}
Polymer foams block IR radiation in two ways.  At infrared frequencies, vibrational modes of the polymers cause significant absorption across the 300 K blackbody spectrum \cite{Smith_1975}. At the same time diffuse scattering (Appendix~\hyperref[appendix:A]{A}) causes a high probability of backscatter and increases the effective path length, further driving up absorption \cite{Amic_1996}. Millimeter wave cameras exploit these effects to reject IR radiation and reduce thermal loading \cite{SO_2020, POLARBEAR, BICEP_windows, Sobrin_2022} on the cryogenic detectors. Since IR radiation must be blocked at the warmest cryostat stages these filters are deployed as the first optical element behind the cryostat window. 

Styrofoam is an expanded closed-cell foam produced by heating predominantly polystyrene combined with fire retardants, blue colorants, and two volatile blowing agents---butane and chloroethane---that vaporize and expand the softened polymer.\footnote{\footnotesize \url{http://www.dupontstyro.co.jp/styrofoam/product/styrofoam.php}}$^{,}$\footnote{\footnotesize \url{http://www.dupontstyro.co.jp/images/sds/SDS01.pdf}} Fire retardants are included for safety as the blowing agents are not fully combusted during fabrication, which leads to molecular outgassing from the media over time\footnote{\footnotesize \url{http://www.dupontstyro.co.jp/styrofoam/notice.html}}, that may contribute to optical variability between samples.

Zotefoam is a closed-cell, cross-linked polyethylene block foam produced by melting polyethylene in a pressurized nitrogen environment and expanded by exposure to the low pressure exterior.\footnote{\footnotesize \url{http://www.Zotefoams.com/who-we-are/3-stage-process/}} Two families are common: the HD foams which are made from a high density polyethelyne (HDPE) substrate, and the LD family is based on a low density polyethelyne (LDPE) substrate. For mm-wave applications, Zotefoams are deskinned to remove a thin residual skin layer leftover from fabrication via a hot wire or similar means (such as cutting, milling, or other). Zotefoam is available in a density range from 15 kg/m$^3$ (LD15) to 110 kg/m${^3}$ (HD110) and is available in black (carbon-loaded) or white (no additives). Zotefoam outgasses over time in vacuum environments and, after pressure cycling, Zotefoam shrinks as the depleted cells provide insufficient pressure support. For this reason, Zotefoam filters must be glued or fastened to the temperature stage. Zotefoam reinflates when pumped to a low vacuum. During operation, the cryogenic conditions freeze out and suppress the pressure from these outgassing molecules. 

\begin{figure}[t]
  \centering
  \begin{subfigure}{0.30\linewidth}
    \begin{overpic}[width=\linewidth]{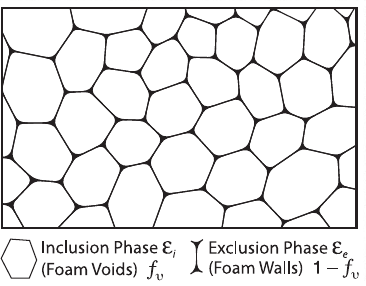}
      \put(43,-15){\textbf{(a)}}
    \end{overpic}
  \end{subfigure}\hfill
  \begin{subfigure}{0.30\linewidth}
    \begin{overpic}[width=\linewidth]{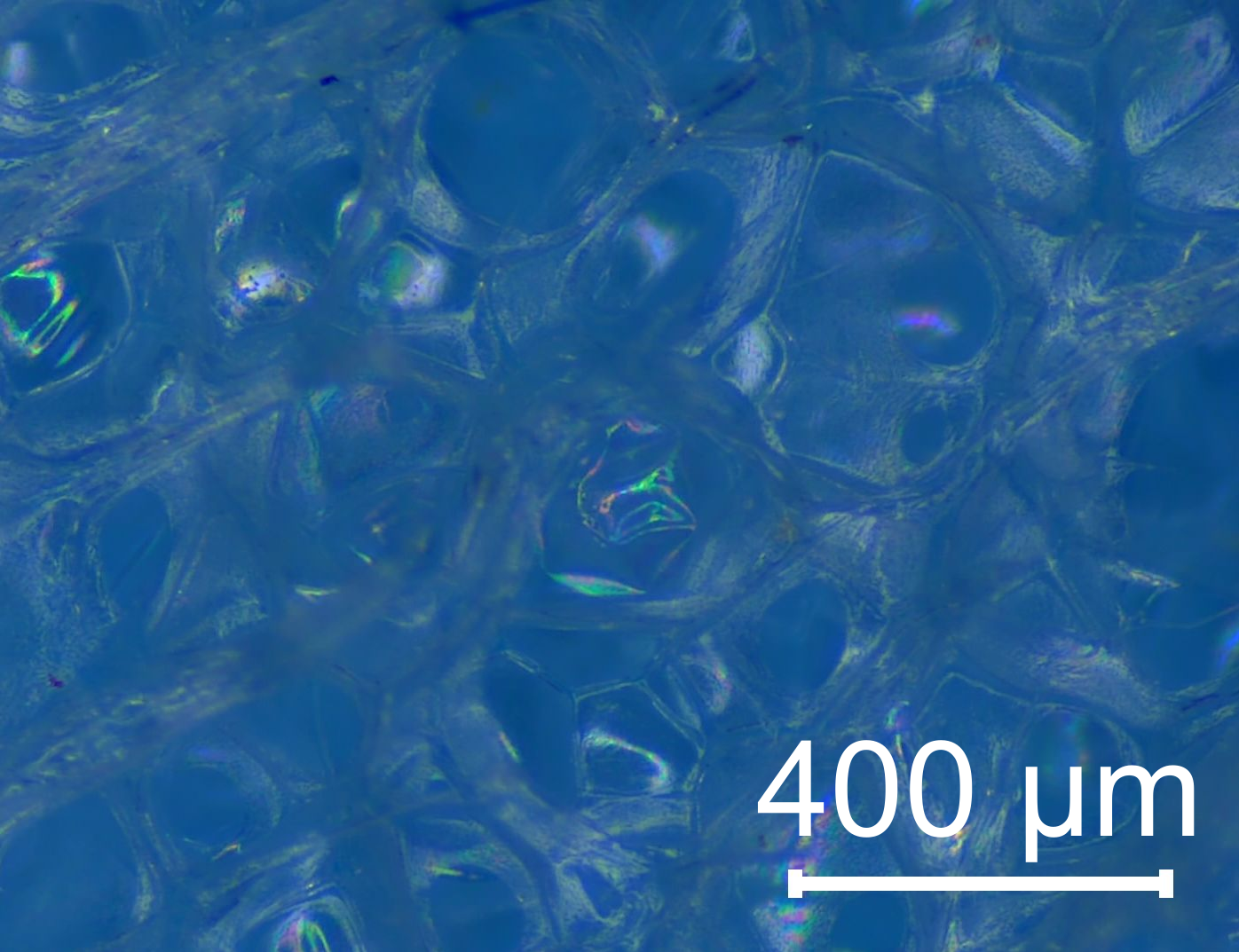}
      \put(43,-15){\textbf{(b)}}
    \end{overpic}
  \end{subfigure}\hfill
  \begin{subfigure}{0.30\linewidth}
    \begin{overpic}[width=\linewidth]{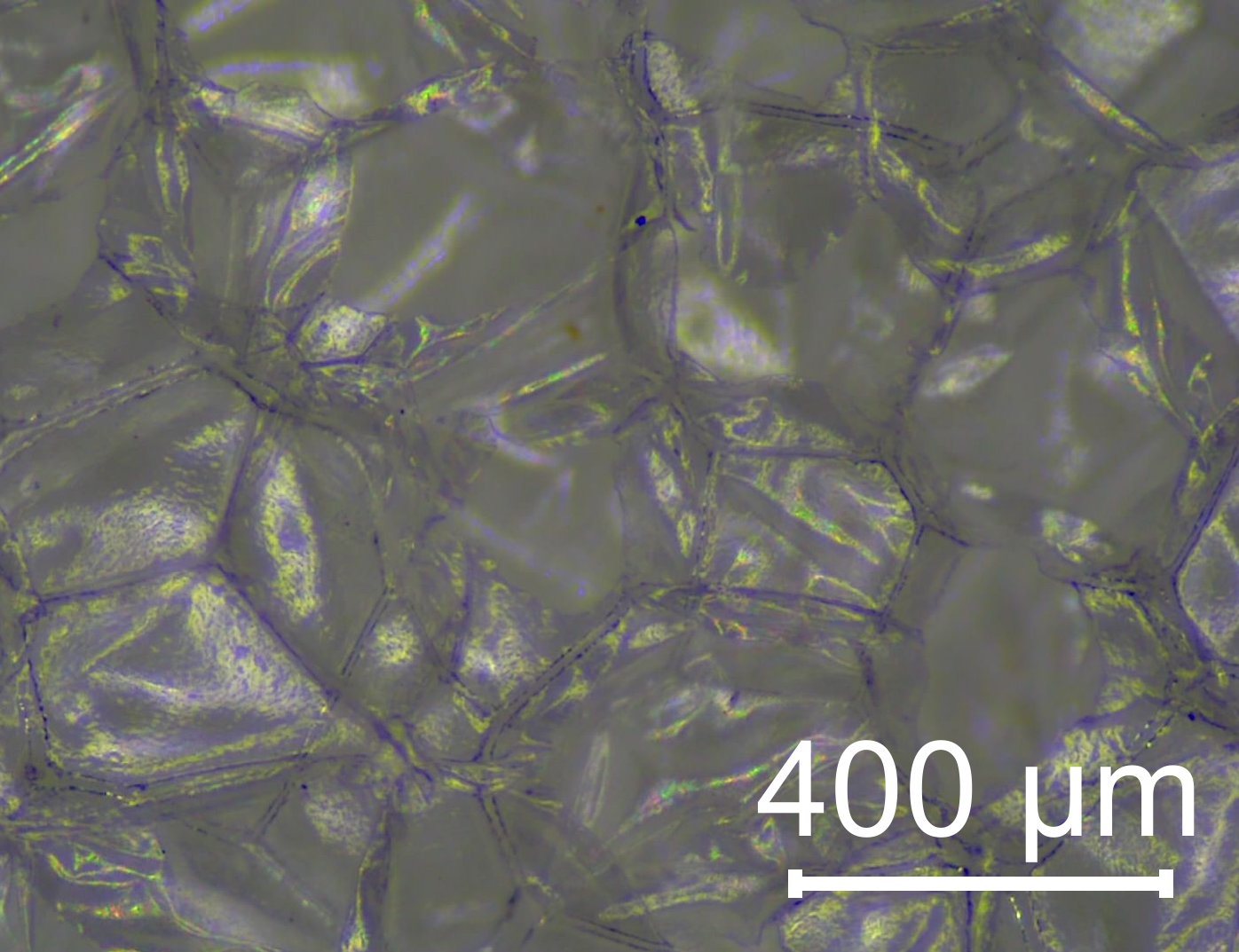}
      \put(43,-15){\textbf{(c)}}
    \end{overpic}
  \end{subfigure}
  \vspace{1em}
  \caption{(a) Sketch of closed-cell foam composed of cells and cell wall intersections, included with permission from \cite{EDW_notes}. Microscope image of Styrofoam (b) and Zotefoam HD30 (c).}
  \label{fig:sketch_images}
\end{figure}

In both foams, nucleation sites expand until a pressure equilibrium is reached. The cell intersections form thin walls, producing the characteristic cell–matrix structure which is shown in illustration and with photographs in Fig.~\ref{fig:sketch_images}. The walls, the plateau borders (where walls intersect), and the cells represent distinct features that are expected to scatter light with differing amplitudes and frequency dependencies. Density inhomogeneity on large scales compared to the cells could lead to additional scattering effects. 

As the cells are randomly oriented with no preferred direction, incident radiation maintains its original polarization state. While scattering phenomena at wide angles can induce polarization (see Appendix~\hyperref[appendix:C]{C}), an emissive or reflective forebaffle blocks radiation at wide angles from entering the telescope in all modern CMB experiments. The resulting optical loading is therefore unpolarized assuming power incident on the forebaffle is uniform. 
Therefore, based upon observational data and theory we believe these material pose minimal risk to polarization measurements.

Styrofoam samples with nominal thickness of 0.15~cm, and 0.6~cm are stacked to create filters. The SO SAT filter stack, as reported in \cite{Day-Weiss_2026}, comprises 22 thin layers between two thick layers with a nominal thickness of 4.5~cm. We retrieved and measured an SO SAT filter, confirming the deployed filters are consistent with other Styrofoam samples discussed in this work that did not undergo cryogenic cycling.

Zotefoam HD30 samples of two nominal thicknesses: 0.32~cm and 2.54~cm are also measured. We measure both single samples and stacked assemblies.
We focus on HD30 due to its historic usage in CMB experiments \cite{BICEP_windows, Sobrin_2022, SO_2020}. Three batches of HD30 from the same manufacturer are compared. We also test less common Zotefoams including  HD110 and HD80; HD60 loaded with carbon inclusions; and LD24 and LD15. All HD30 and denser foams are known to possess the necessary structural integrity for cryogenic applications. We conducted three rounds of LD24 vacuum testing, confirming it remains mechanically robust at pressure near zero bar. Cryogenic LD24 testing is underway within an SO SAT, while LD15 remains vacuum and cryogenically untested. Two batches of LD15 and LD24 are compared, while we declined to source secondary batches of the denser Zotefoams due to excess attenuation.

\section{Spectrometer Measurement}
\label{sec:measurement}
Data are taken using a TeraFlash pro, a pulsed laser Fourier transform time domain terahertz spectrometer (TD-THz) manufactured by Toptica Photonics, which measures broadband spectra with a peak dynamic range of ${\gtrsim}45$~dB at a rate of seconds for each measurement.\footnote{\url{www.toptica.com/fileadmin/Editors_English/11_brochures_datasheets/03_Short_Info/toptica-TeraFlash_pro-short-info.pdf}}
The instrument employs asynchronous optical sampling (ASOPS) with two mode-locked femtosecond fiber lasers of slightly detuned repetition rates, allowing the full temporal waveform to be sampled without mechanical delay stages. The recorded photocurrent pulse has a tunable duration between 5 and 2000~ps, with amplitudes ranging from 0.01 to 1000~nA, and is sampled with a fine temporal resolution of 0.05~ps yielding spectra from 5~GHz to 10~THz. Analysis was performed using \texttt{TeraToptica} \cite{TeraToptica}, an open-source Python package for the Toptica TeraFlash and TeraScan instruments.

\begin{figure}[t]
    \centering
    \includegraphics[width=\linewidth]{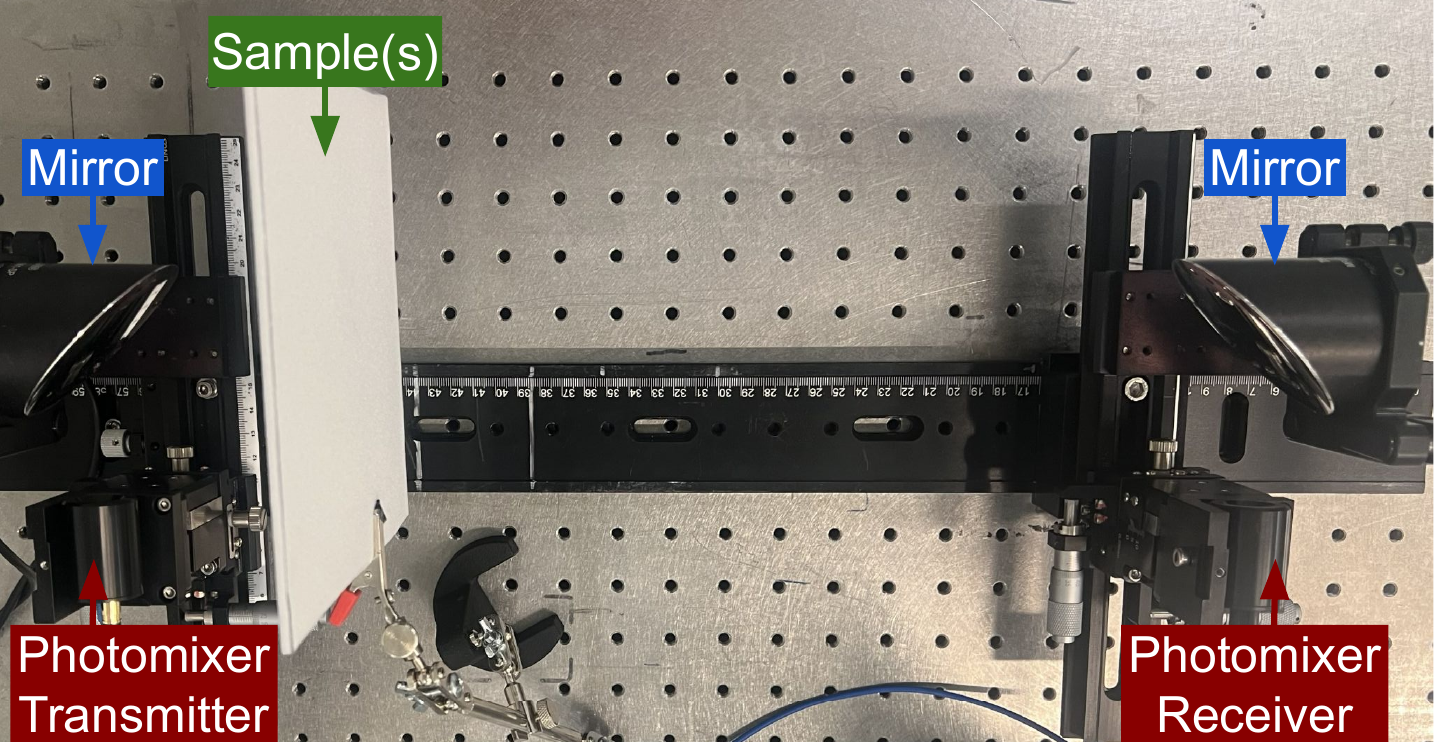}
    \caption{One nominally 0.32~cm thick HD30 sample mounted in the collimated TeraFlash beam. Varying the locations of sample(s) within the beam yields no detectable difference in our transmittance measurement or model results.}
    \label{fig:setup}
\end{figure}

Our experimental design uses two 5.08~cm off-axis parabolic mirrors to collimate and focus the spherical wave. Emitting and recieving photomixers are aligned to maximize the time-domain pulse photocurrent into a coherent plane wave normal to the samples, shown in Fig.~\ref{fig:setup}. The photomixer photocurrent drifts upward after power-on, with the largest increase occurring within the first hour. As our calibration relies on the ratio of the sample to free-space reference spectra, this warm-up transient can bias the inferred transmittance if the two measurements are performed at different points in the drift. We therefore allow the instrument to stabilize before acquiring data.


\subsection{Time Domain}
\label{sec:td}
\begin{figure}[ht]
    \centering
    \includegraphics[width=\linewidth]{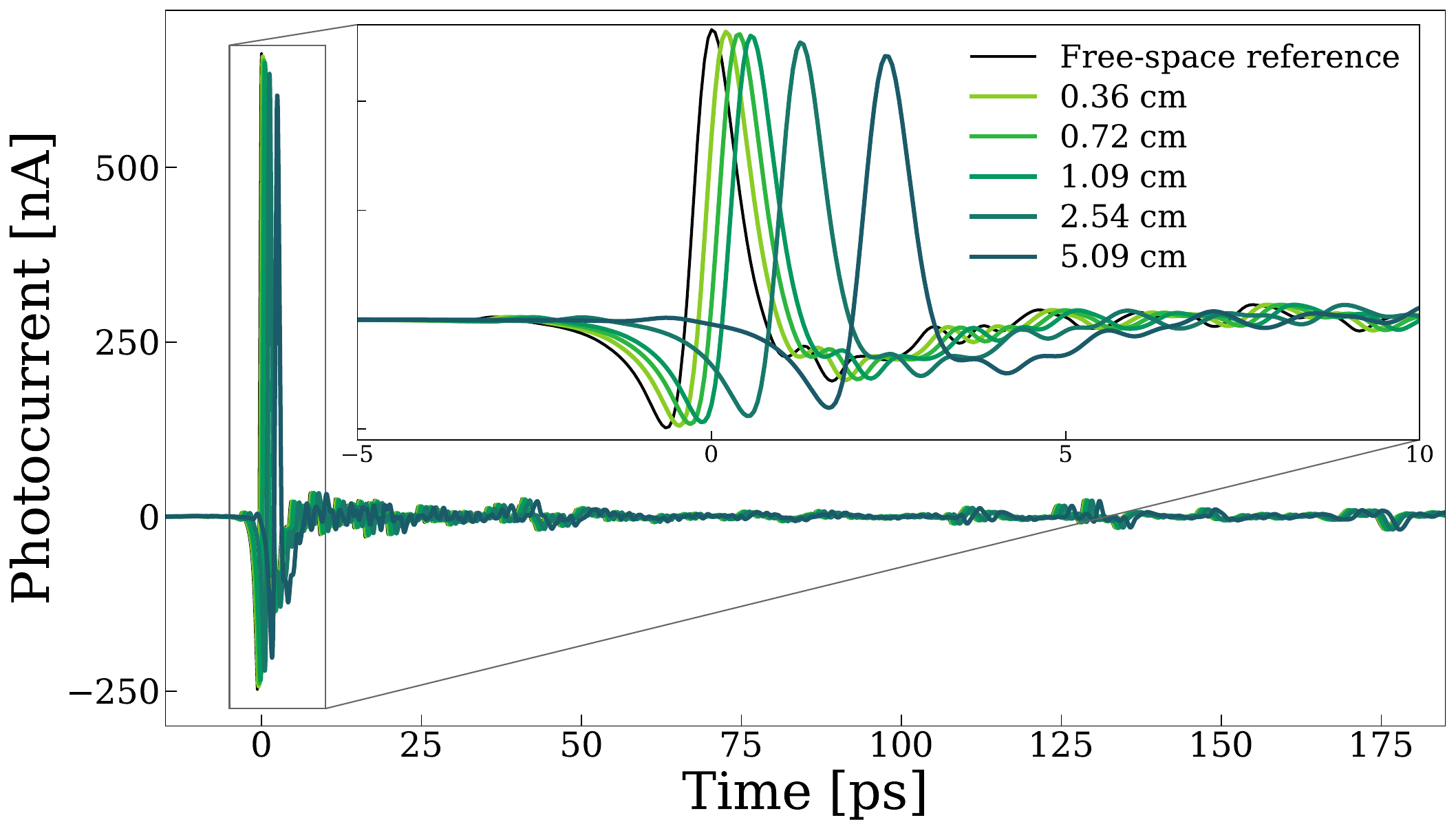}
    \caption{
    Time-domain photocurrent amplitude for various thicknesses of HD30 assemblies sourced from the same batch. The free-space reference is shown in black, with $t=0$ chosen to be the peak of the reference pulse. Increasing assembly thickness yields a proportional temporal shift and reduced amplitude. The zoomed view highlights the early-time response.}
    \label{fig:zote_time_domain}
\end{figure}

Because the broadband terahertz pulse undergoes chromatic dispersion as it propagates through a medium, different wavelengths arrive at different times. Measuring the full temporal waveform therefore captures both the group delay and phase shift introduced by the sample. A Fourier transform of the time-domain signal recovers the frequency-domain transmittance spectrum with resolution $\Delta f = 1/T$, where $T$ is the chosen pulse duration. To mitigate reflections internal to the emitter and receiver, the time-domain approach allows temporal gating to eliminate these systematic interference effects \cite{Toptica_2010, Roggenbuck_2010}, leading to cleaner and readily interpretable spectra. The rapid acquisition of time-domain pulses shown in Fig.~\ref{fig:zote_time_domain} makes practical the characterization of numerous polymer foam materials. 

\subsection{Frequency Domain}
Frequency-domain spectra with 5 GHz (1 GHz) resolution are obtained using a Fourier transform of the 200~ps (1000 ps) time-domain pulse, equivalent to a path length of 6~cm (30~cm) in free space. In this work, we use a 200~ps time-gated pulse to eliminate reflections internal to the silicon lenslet encased photomixers that occur shortly outside this window resulting in more easily interpretable spectra.
The Fourier transform is implemented as a Blackman window and fast Fourier transform (FFT), computed automatically using the vendor supplied software.  
The coherent detection scheme of the TeraFlash pro system measures both the phase and amplitude information of the resulting spectra. 
A number of measurements, 100 in this work, are obtained sequentially and averaged to reduce the photon noise with the mean phase and amplitude of the chosen number returned after the measurements are complete. 
Supply-line standing waves arising from the antenna are not mitigated by time gating and produce fringes below ${\sim}150$~GHz \cite{stanze_2011}. Meanwhile, diffuse scattering arising in the samples along with low photocurrent signal to noise effectively limits measurement to ${\sim}2$~THz. The spectra are binned with a minimal 15 GHz resolution (n=3) to obtain the local systematic uncertainty. We also consider the errors from the uncertainty on the measured width, the instrumental SNR via the dark current, and statistical variation between samples within a batch. The statistical uncertainty within a batch dominates in all cases.

\begin{figure}[h]
    \centering
    \includegraphics[width=\linewidth]{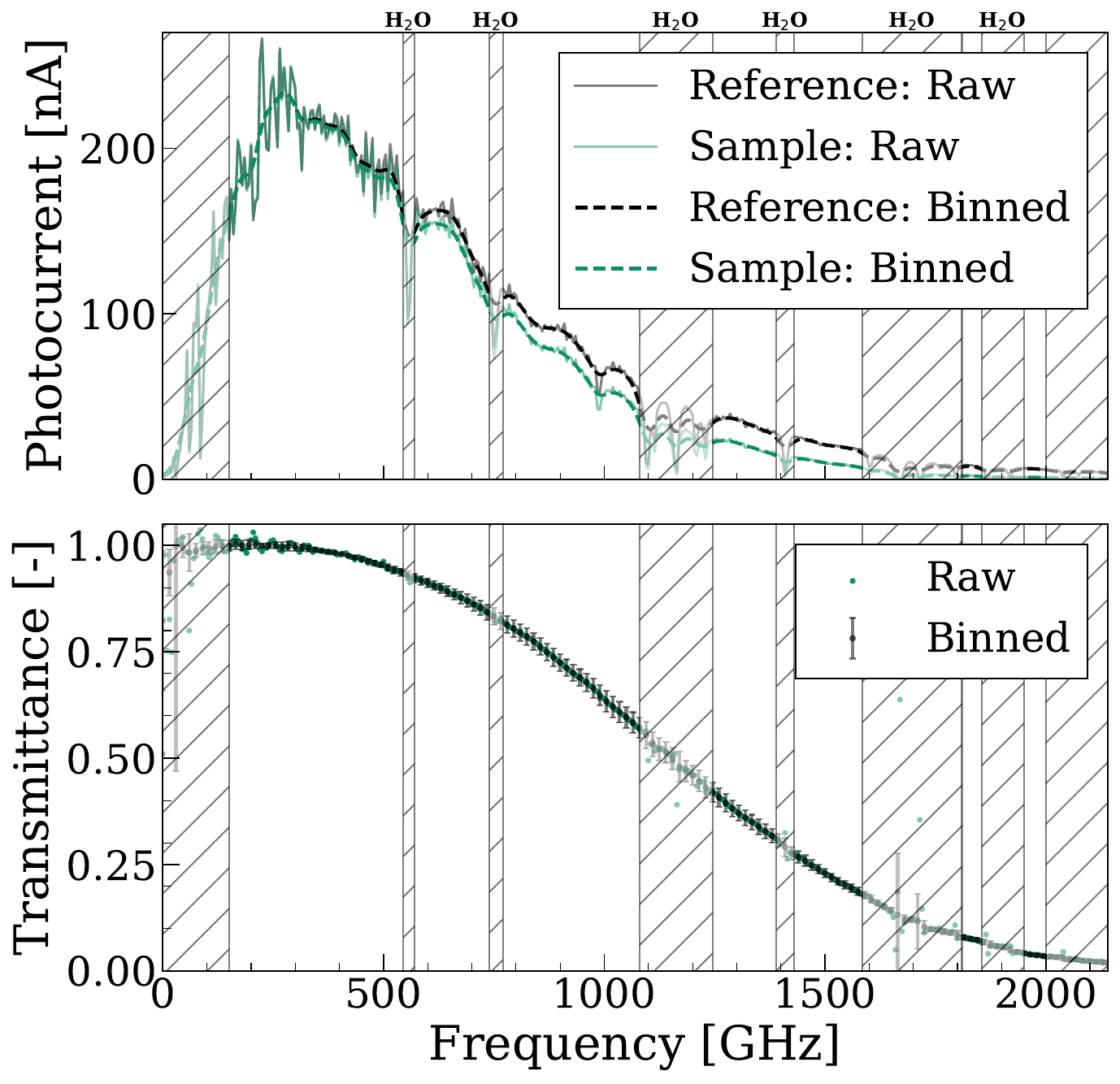}
    \caption{(Upper) Raw versus 15 GHz binned photocurrents for a 2.54~cm Zotefoam HD30 sample and free-space reference. (Lower) The corresponding transmittance spectrum with uncertainties. Uncertainties are used to weight the fitting. Binning is employed to mitigate residual low frequency systematic fringes. Data below 150~GHz and above 2,000~GHz where the low photocurrent limits accuracy are excluded from our analysis. Intermediate regions that are contaminated by waterlines are masked as well.}
    \label{fig:ut}
\end{figure}

We calibrate our transmittance spectra by taking a reference measurement through free space. We then divide the spectrum through the sample by this reference and square to convert from amplitude to power and obtain the transmittance of the sample.
These transmittance spectra exhibit contamination from atmospheric absorption features due to molecular absorption (principally H$_2$O, with additional contributions from O$_2$ and CO) shown in Fig.~\ref{fig:ut}. Only the H$_2$O transitions produce features of sufficient strength to require exclusion. To prevent artifacts associated with smoothing of the transmittance spectra which broadens these features, we conservatively mask data over frequency intervals wider than the smoothing kernel size from each line center listed in Appendix~\hyperref[appendix:E]{E}.

The measured spectral phase encodes the chromatic dispersion of the sample, arising from the frequency dependence of the effective refractive index. To extract $n_{\rm eff}(\lambda)$, we first compute the change in the optical path length $\Delta L(\lambda)$ from the phase difference between the sample spectrum and a free-space reference spectrum $\Delta \varphi(\lambda)$. The effective refractive index for a sample of thickness $w$ is then obtained via

\begin{equation}
    n_{\rm eff}(\lambda) = 1 + \frac{\Delta L(\lambda)}{w} = 1 + \frac{\lambda\Delta \varphi(\lambda)}{2\pi w}
    \label{eq:n_eff}
\end{equation}

While the time delay of pulse also encodes the refractive index in principle, waterlines that positively bias the time delay would result in systematic overestimation. After masking atmospheric waterlines in the phase difference, we weight the data using the photon-noise uncertainty to obtain the effective refractive index for each sample. We then equally weight batches to obtain $\langle n_\text{eff} \rangle = 1.016\pm 0.002$ for Zotefoam HD30 and $\langle n_\text{eff} \rangle = 1.022\pm 0.001$ for Styrofoam. Our Styrofoam value is consistent with the 1.017 - 1.022 range reported in \cite{Dietlein_2008} for similar formulations.

\section{Transmittance Model}
\label{sec:model}

To extract the mm-wave optical properties, we fit the data with a radiative transfer model that incorporates dielectric absorption, Mie scattering, and Rayleigh scattering. Reflections are neglected, since for these foams with $n_\mathrm{eff} \sim 1.01$ the Fresnel reflection coefficient is $R \sim 10^{-4}$. Rayleigh scattering considers only the dipole polarizability of the scattering center and is appropriate for scatterers small compared to the wavelength. Mie scattering accounts for the phase shift accumulated as a wavefront traverses scatterers of size comparable to the wavelength.

The characteristic cell–matrix structure of the material (Fig.~\ref{fig:sketch_images}) motivates a transmittance model including both Rayleigh and Mie scattering components as the hundreds-of-microns cells are comparable to the wavelengths considered here while the few-micron-thick cell walls are firmly in the Rayleigh regime. There are, therefore, two distinct scattering varieties. The physical length scale and frequency dependence of each pathway determines their hierarchy. At longer wavelengths, scattering is dominated by the cells inducing Mie scattering. At shorter wavelengths, the steeper Rayleigh scattering efficiency and larger refractive-index contrast between the polymer shells and the matrix voids cause Rayleigh scattering from the cell walls to dominate the total scattering.

The transmittance including these terms can be written as
\begin{equation}
\small T(f) = T_{\rm Mie}(f)\,T_{\rm Ray}(f)\,T_\delta(f) = \exp[-w(\alpha_M(f)+\alpha_R(f)+\alpha_\delta(f))]
\end{equation}
where $T_{Mie}$, $T_{\rm Ray}$, and $T_\delta$ represent the transmittance for a Mie, Rayleigh, and absorption model; and $\alpha_M$, $\alpha_R$, and $\alpha_\delta$ are the respective attenuation components. The filter transmittance can be normalized by rescaling to a consistent filter thickness $w$, 5~cm in this work.

The absorption for Mie scattering is typically written in terms of a size parameter related to the cell size as well as the cell and matrix refractive indices (see Eq.~{\ref{eq:qmie}). However, since these properties are not known with certainty, we introduce $K$ parameters \cite{Shachar_2020} to encode these properties and constrain this parameter by fitting the measured spectra. To ensure the attenuation has the proper units, we fit a prefactor $C$ with units of inverse length (related to the effective number of scattering events per unit length). With these considerations, the Mie scattering attenuation is defined as
\begin{equation}
    \alpha_M(f) \equiv C\left[2 - 4\frac{\sin(f/K_{\mathrm{Mie}})}{f/K_{\mathrm{Mie}}} + 4\frac{1-\cos(f/K_{\mathrm{Mie}})}{(f/K_{\mathrm{Mie}})^2}\right]
    \label{eq:alpha_M}
\end{equation}
where $f$ is the frequency, and $K_{\rm Mie}$ encodes material properties.

Similarly, we define an expression for Rayleigh scattering as
\begin{equation}
   \alpha_R(f) \equiv C(f/K_{\mathrm{Ray}})^4
   \label{eq:alpha_R}
\end{equation}
where we similarly introduce $C$ to ensure all $K$ parameters have identical units of frequency. We derive expressions for $K_{\textrm{Mie}}$ and $K_{\textrm{Ray}}$ in terms of the material properties in \S~\ref{sec:cst}. More information about these processes is provided in Appendix~\hyperref[appendix:A]{A}.

Comparatively, the intrinsic absorption is determined predominantly by the bulk dielectric loss with the cell properties bearing minimal impact (\S\ref{sec:bp}). It is defined as

\begin{equation}
    \alpha_\delta(f) \equiv \frac{2\pi n_{\rm eff}\tan\delta}{c}f
\end{equation}
where $n_{\rm eff}$ is the effective refractive index, $\tan(\delta)$ is the dielectric loss tangent, and $c$ is the speed of light. The polymer material, HDPE for Zotefoam and Polystyrene for Styrofoam, is the continuous solid phase of the foam that defines the imaginary component of permittivity by dominating the absorption.

\subsection{Coupled Secondary Scattering}
\label{sec:scat2}

Both Rayleigh and Mie scattering are leading approximations of more complex physical processes. As the wavelength approaches the cell size, proper treatment must consider multiple scattering events and the tendency of the cell walls to act as dielectric wave guides. To maintain accuracy into the terahertz regime, we model secondary scattering where scattered light is rescattered within the media by successive interactions. Although the probability of rescattering on-axis is vanishing, the detector integrates over a nonzero solid angle, leading to a contribution from successive scatters. We find that our fits have large residuals beyond ${\sim}$500~GHz unless secondary scattering terms are included. We therefore update the transmittance model to

\begin{equation}
T = \exp[-w(\alpha_\delta+\alpha_R+\alpha_M- S_2)].
\end{equation}
\noindent The secondary scattering contribution $S_2$ is parameterized as
\begin{equation}
S_2 = \big(a_{RR} \alpha_R^2+ 2\,a_{RM}\,\alpha_R\alpha_M + a_{MM} \alpha_M^2\big)/C
\label{eq:scat2}
\end{equation}
where $a_{RR}$, $a_{RM}$, $a_{MM}$ are dimensionless coupling efficiencies between 0 and 1 corresponding to the two-scatter pathways. The Rayleigh--Mie channel appears with a factor of 2 since the two-scatter process is independent of the order of Rayleigh and Mie scattering events. Each term is divided by $C$ to ensure dimensionality. Scattering and absorption are treated as independent processes, a well-justified assumption for minimally-absorptive media \cite{vdH}.

We find evidence for a  coupling between Mie scattering and Rayleigh scattering that adds power at high frequency.  We investigated other two scatter processes including the Rayleigh-Rayleigh channel, which was not detected. The Mie-Mie channel is formally degenerate with Rayleigh scattering to leading order; we tested this degeneracy by building models using only expansions of higher order Mie scattering with no Rayleigh component ($\alpha_R = 0, a_{RR} \neq 0$, and introducing $a_{RRR})$ which fail to empirically describe the transmission below ${\sim}300$ GHz. For this reason we set $a_{RR} = a_{MM} = 0$. This result agrees with theoretical expectations. Only the Rayleigh-Mie channel is detected, owing to the inherent coupling of two distinct classes of scatterers (low-density voids and the surrounding polymer shells). The mathematics for higher order scattering corrections are also developed in Appendix~\hyperref[appendix:D]{D}.

With these simplifications, the transmittance model is
\begin{equation}
    T = \exp[-w(\alpha_\delta+\alpha_R+\alpha_M- 2a_{RM}\alpha_R\alpha_M/C)]. \label{eq:T}
\end{equation}

To obtain parameter constraints and uncertainties, we run a nested sampling routine using \texttt{Dynesty} \cite{dynesty} to fit $C$, $K_{\rm Ray}$, $K_{\rm Mie}$, $\tan\delta$, and $a_{RM}$ in Eq.~\ref{eq:T}.
Nested sampling was employed for robustness against local optima \cite{Feroz_2008}.
Coupling efficiencies use the physically motivated priors of [0, 1] for convergence; with all other parameters insensitive to a wide range of tested priors. We achieve broad convergence from 150~GHz to 2~THz with exceptions of HD80 and HD110, which have higher densities leading to diffuse scattering below 2~THz.

With the exception of carbon-loaded samples, a dielectric loss tangent is not detected in any samples as the in-band absorption is subdominant to the scattering losses. However, as dielectric loss results in relatively more optical loading than scattering for same percentage loss due to blackbody emission from the around 200 K foams, the optical loading from emission is non-negligible. We therefore estimate a loss tangent from material properties (see \S\ref{sec:bp}) to predict the field-performance.


\subsection{Styrofoam Fit Parameters}
We measure Styrofoam and show a single model fit to the combined dataset seen in Fig.~\hyperref[fig:spectra]{\ref*{fig:spectra}, left}. Styrofoam samples were sourced directly from a SO 220/280 GHz telescope, confirming deployed filters were consistent with other samples from the same batch that did not undergo cryogenic cycling. We show the parameters of both the combined batch fit and individual assembly fits in Table~\ref{tab:params}. 
For a typical 5~cm thick filter at 280~GHz, Styrofoam yields $90.1\pm0.2\%$ transmittance. 

A subset of Styrofoam data were not well fit by our model. These behavior was repeatable in Styrofoam, but similar behavior was not found in any other foams. We compared the samples that were well described the model and those that were not, finding that the scattering parameters were consistent between these two subsets while the samples poorly described by our model implied unphysically low loss tangents. This inconsistency is likely evidence of inhomogeneities at the scale of the illuminating mode as repeated measurements of different locations on the same samples yield inconsistent results. Such inhomogeneities could arise from clusters of the copper phthalocyanine pigment which have semiconductor-like losses \cite{Abkowitz_1972}, structural damage to the media in sample preparation, or poorly mixed blowing agents yielding variable optical properties. Due to these imohogeneities, we model samples individually and report the largest loss tangent upper bound in Table~\ref{tab:params}. These inhomogeneities imply the optical properties of styrofoam should be carefully measured before incorporation as an optical element.

Styrofoam IR filters were presented in Choi \textit{et al.}~\cite{Choi_2013} which defined an approximate Styrofoam transmittance function at 200 GHz based on the number of layers. As the mm-wave transmittance depends on thickness, we reformulate their approximation  to depend on the width $w$ in millimeters as $T\gtrsim 0.999^w$ for comparability. We find $T= (0.99894\pm0.00004)^w$ at 200 GHz, and therefore believe our Styrofoam batch to be optically similar to the samples studied in \cite{Choi_2013}.

\subsection{Zotefoam HD30 Fit Parameters}
\label{sec:zf_model}

\begin{figure*}[t]
    \centering
    \includegraphics[width=\linewidth]{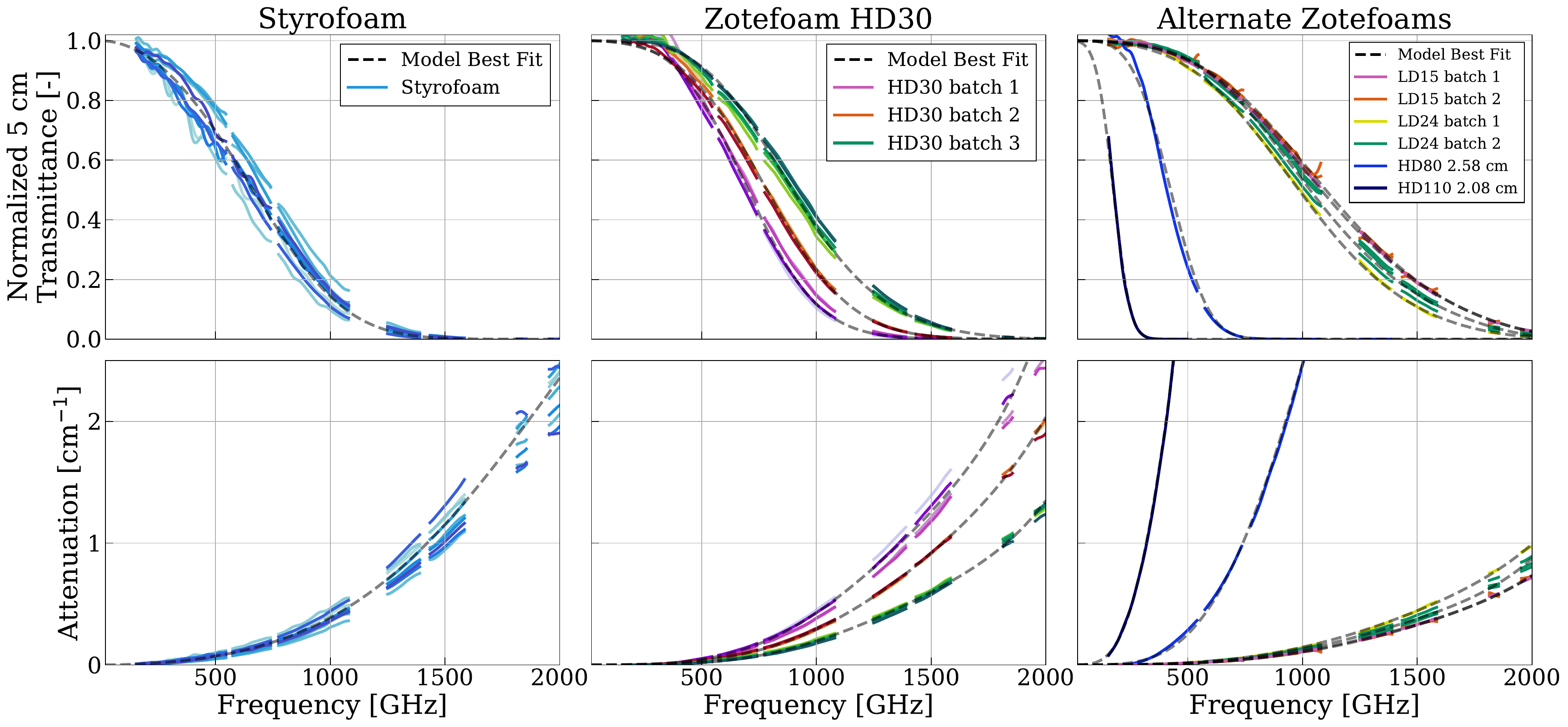}
    \caption{Binned transmittance (upper) spectra normalized to a 5~cm thick filter and attenuation (lower) spectra for Styrofoam and Zotefoams. The dashed line is the secondary scattering model best fit for each batch. Sample thicknesses range from 0.1 to 5.1 cm, darker color indicates thicker samples. Convergence of the model fit from 150~GHz to 2~THz yields narrow parameter and mm-wave performance constraints. The spectra within a given HD30 batch is consistent while performance varies between batches, which we call batch-to-batch variability. Similar attenuation between assemblies of different numbers of layers confirms Ruze scattering from surface errors is subdominant to Rayleigh and Mie scattering. For alternate Zotefoams, the attenuation decreases with lower density in all but one case, where batch-to-batch variability results in one LD15 batch possessing higher attenuation than an LD24 batch.}
    \label{fig:spectra}
\end{figure*}
We measured three batches of Zotefoam HD30, each sourced separately from the same manufacturer, and shown a single model fit to each combined dataset in Fig.~\hyperref[fig:spectra]{\ref*{fig:spectra}, center}. While each batch is well described by its respective model, variability between batches occurs. This is likely intrinsic to foam manufacturing differences relevant for its optical properties but irrelevant for its typical uses. 
For a typical 5~cm thick filter at 280~GHz, the highest- and lowest-transmittance batches yield $98.88\pm0.04\%$ and $96.7\pm0.2\%$, respectively—a batch-to-batch spread of ${>}2\%$ that motivates optical characterization on a batch-by-batch basis before deployment.
Increasing individual sheet thickness beyond what is required for IR opacity decreases the in-band transmittance (Eq.~\ref{eq:T}) without further attenuating IR (Eq.~\ref{eq:MLE}). We therefore do not consider 5.08~cm or 7.62~cm thick Zotefoams.\footnote{We additionally identified these formulations are created by bonding 2.54~cm samples together, leaving the skin layer unremovable and likely yielding scattering.}


\begin{table*}[t]
\centering
\renewcommand{\arraystretch}{1.4}
\begin{tabular}{lcccccc}
\hline\hline
Sample 
& $\tan\delta^*$
& $C [\mathrm{cm}^{-1}]$ 
& $K_{\mathrm{Ray}} [\mathrm{GHz}]$ 
& $K_{\mathrm{Mie}} [\mathrm{GHz}]$ 
& $a_{\mathrm{RM}}$
& $n_{\mathrm{eff}}$ \\
\hline
Styrofoam batch & $<3\times10^{-4}$ & $3^{+1}_{-1}$ & $2000^{+200}_{-300}$ & $2300^{+400}_{-700}$ & $0.7^{+0.2}_{-0.3}$ & $1.020(3)$ \\
HD110 & -- & $0.023$ & $112$ & $148$ & $0.103$ & $1.0636(9)$ \\
HD80 & -- & $4.9\times10^{-4}$ & $101$ & $451$ & $0.129$ & $1.0400(7)$ \\
HD30 \\[-0.5em]
\qquad batch 1 & $<1.3\times10^{-5}$ & $0.017^{+0.004}_{-0.004}$ & $410^{+20}_{-30}$ & $488^{+9}_{-8}$ & $0.114^{+0.001}_{-0.001}$ & $1.019(2)$ \\
\qquad batch 2 & $<1.5\times10^{-5}$ & $0.035^{+0.003}_{-0.003}$ & $560^{+10}_{-10}$ & $533^{+6}_{-6}$ & $0.1042^{+0.0008}_{-0.0008}$ & $1.014(1)$ \\
\qquad batch 3 & $<5.6\times10^{-6}$ & $0.0025^{+0.0008}_{-0.0009}$ & $300^{+20}_{-30}$ & $500^{+4}_{-4}$ & $0.1139^{+0.0007}_{-0.0007}$ & $1.0147(6)$ \\
LD24 \\[-0.5em]
\qquad batch 1 & -- & $0.0206$ & $593$ & $522$ & $0.1034$ & $1.0117(1)$ \\
\qquad batch 2 & $<1.5\times10^{-5}$ & $0.004^{+0.001}_{-0.002}$ & $380^{+30}_{-50}$ & $494^{+8}_{-7}$ & $0.112^{+0.001}_{-0.001}$ & $1.0123(2)$ \\
LD15 \\[-0.5em]
\qquad batch 1 & -- & $0.0138^{+0.0003}_{-0.0003}$ & $567^{+4}_{-4}$ & $514^{+2}_{-1}$ & $0.1072^{+0.0003}_{-0.0003}$ & $1.0090(1)$ \\
\qquad batch 2 & $<1.1\times10^{-5}$ & $0.015^{+0.002}_{-0.002}$ & $600^{+30}_{-30}$ & $532^{+10}_{-9}$ & $0.101^{+0.002}_{-0.002}$ & $1.0075(7)$ \\
\multicolumn{7}{l}{\footnotesize$*$ Loss tangent not detected, the 95th percentiles of the posterior are listed.} \\
\hline\hline
\end{tabular}
\caption{Table of mean batch fit parameters and $1\sigma$ uncertainties along with the effective refractive indices for the polymer foam samples analyzed. Styrofoam samples with unphysical loss tangents were exluded from fitting; uncertainties are not reported for single sample batches as dominate uncertainty, variation between samples, is unknown. All foams with multiple batches have batch-to-batch variability. We obtain the effective refractive index of each batch via the phase delay (Eq.~\ref{eq:n_eff}), excluding the diffuse scattering regime of HD80 and HD110.}
\label{tab:params}
\end{table*}

\subsection{Alternate Zotefoam Materials}
We compare measurements of LD15, LD24, HD30, HD80, and HD110 in Fig.~\hyperref[fig:spectra]{\ref*{fig:spectra}, right}. We find that higher (lower) density Zotefoams exhibit steeper (shallower) attenuations as expected with thicker (thinner) cell walls and similarly sized cells (see Appendix~\hyperref[appendix:B]{B}). For a typical 5~cm thick filter at 280~GHz, the first batches of LD24 and LD15 yield $98.1\%$ and $98.6\%$ transmittance, respectively. These batches are single samples, and the dominant uncertainty, from sample variations, is therefore unknown. The second LD24 batch exhibits improved performance of $99.02\pm0.07\%$, while the second LD15 batch exhibits consistent performance of $98.57\pm0.08\%$ for the same width and frequency band. 

Excess attenuation and diffuse scattering arises in HD80 and HD110 which test the limits of our model and instrument, shown in Fig.~\ref{fig:phase}. Black HD60 samples were excluded from further analysis after we found excess in-band scattering and absorptive losses (e.g., $T \leq 97.3\%$ (83.2\%) for 5~cm in a 90 (220)~GHz band) and degenerate parameters requiring more sophisticated modeling, both resulting from the carbon inclusions. We discuss the optical implications of these Zotefoam formulations and make recommendations for mm-wave experiments in \S\ref{sec:discussion}.

\begin{figure}[h!]
  \centering
  \includegraphics[width=\linewidth]{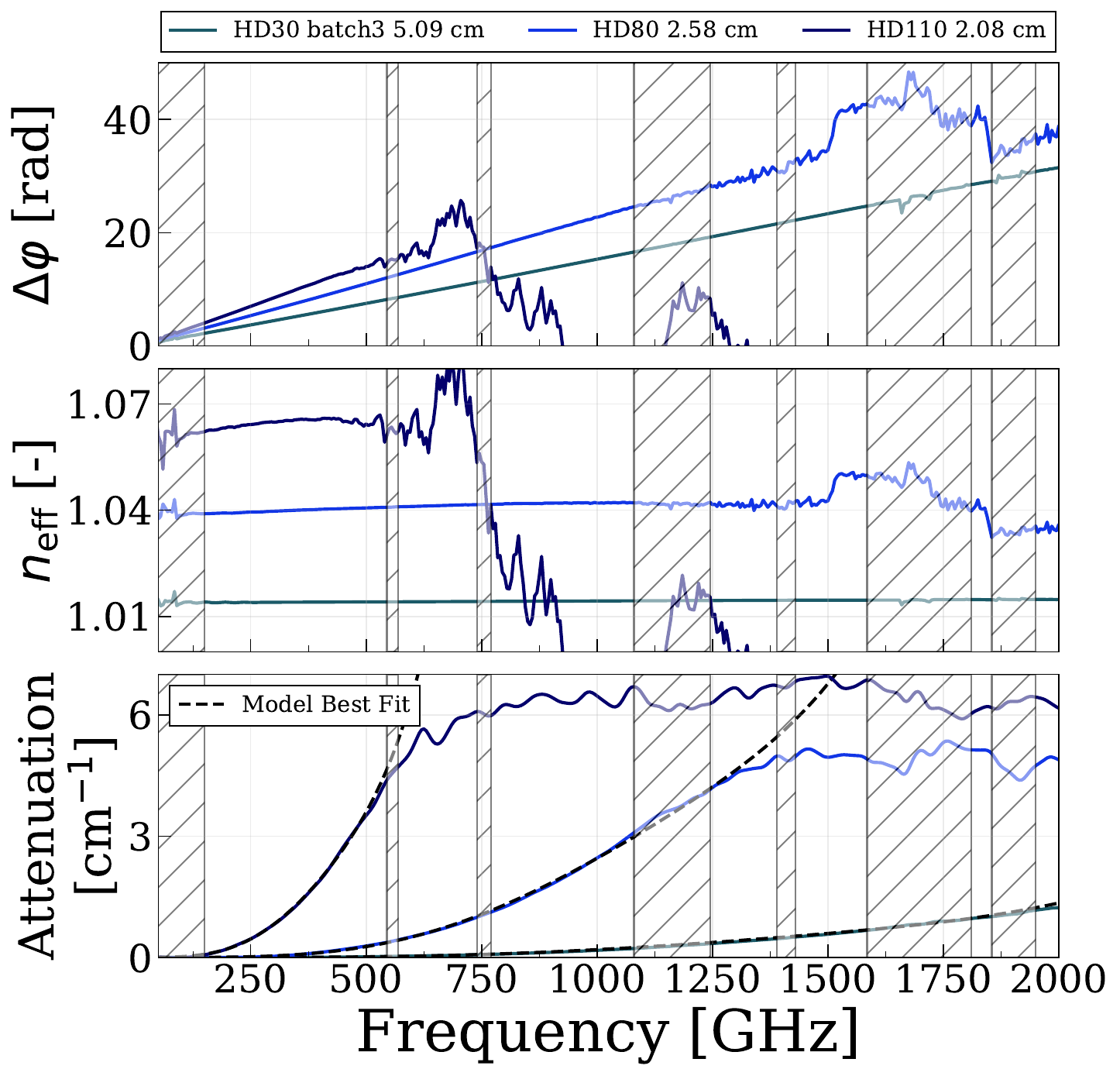}
  \caption{HD80 and HD110 reach the diffuse scattering regime which yields phase errors as the decoherent phase information of scattered photons is recorded, impacting the measured effective refractive index. Diffuse scattering also results in an asymptotic attenuation limit which is not considered by our model.}
  \label{fig:phase}
  \vspace{-1em}
\end{figure}



\section{Physical Properties}
\label{sec:ep}

In this section, we discuss the physical properties of these materials and bridge the physical and optical properties. In \S\ref{sec:bp}, we estimate the fractional matrix volume $f_m$ in two manners, which show consistent results, and state the resulting dielectric loss tangent estimates. In \S\ref{sec:cst}, we estimate the cell number densities and cell wall thicknesses in order to compare our fit results to scattering theory.

\subsection{Bulk properties}
\label{sec:bp}
The refractive indices of the cell voids and polymer match those of the bulk materials.  For HDPE and polystrene $n_{\rm HDPE} \approx 1.54$ and $n_{\rm ps} \approx 1.59$ \cite{James_1996, ISO489, Goldsmith1998}, respectively, and we approximate the index of the voids as one.

Starting with the measurements of the refractive indices presented in \S\ref{sec:td}, we use the Maxwell–Garnett (MG) effective-medium approximation to find the fractional polymer volume $f_p$ in the dilute limit ($f_p \ll 1$) \cite{bohren_huffman, markel_2016},
\begin{equation}
\frac{n_{\mathrm{eff}}^{2} - n_m^{2}}{n_{\mathrm{eff}}^2 + 2 n_m^2}
= f_p \frac{n_p^{2} - n_m^2}{n_p^2 + 2 n_m^2}.
\end{equation}
We then find $f_p~\approx~0.03$ for Zotefoam and $f_p~\approx~0.04$ for Styrofoam where fractional matrix volume $f_m + f_p = 1$.

We compare these results to an estimate derived from the known density,
\begin{equation}
    \rho_\text{eff} = f_p\rho_p + f_m \rho_m \label{eq:rhos}
\end{equation}
adopting densities $30$, ${\geq}25$, $1050$, $950$, and $1.25$ kg/m$^3$ for Zotefoam HD30, Styrofoam, polystyrene, HDPE, and the matrix (pure nitrogen) respectively \cite{HD30, DuPont_StyroaceII, rho_ps, HDPE}. This yields $f_p \approx 0.03$ for Zotefoam, and $f_p \gtrsim 0.02$ for Styrofoam, in reasoanble agreement, demonstrating the robustness of the MG framework and the inferred $f_m$ despite uncertain material properties.

Using $f_p$ from Eq.~\ref{eq:rhos} and $n_{\rm eff}$ from Table~\ref{tab:params}, we estimate the effective dielectric loss tangent. Dielectric loss enters as a small imaginary part of the polymer permittivity, $\tan\delta \equiv
\varepsilon''/\varepsilon'$ with $\varepsilon = n^2$. Treating the polymer permittivity as complex and expanding the MG relation to first order in loss, the effective loss tangent of polymer shells embedded in a lossless void medium is
\begin{equation}
\tan\delta_{\rm eff} = \frac{n_p^2\,(n_{\rm eff}^2 + 2n_m^2)^2}
{n_{\rm eff}^2\,(n_p^2 + 2n_m^2)^2}\, f_p \tan\delta_p. \label{eq:eff_lt}
\end{equation}

We report the resulting fractional polymer volumes and loss tangents from Eqs.~\ref{eq:rhos} and \ref{eq:eff_lt}, respectively, in Table~\ref{tab:eff_lt}. For LD15 and LD24, we use $\rho_{\rm LDPE}~\approx~925$~kg/m$^3$.\footnote{\scriptsize{\url{https://lairdplastics.com/resources/lowdensity-polyethylene-ldpe-complete-technical-guide/}}}

\begin{table}[h]
    \centering
    \small
    \setlength{\tabcolsep}{4pt}
    \begin{tabular}{cc|ccc}
    \hline\hline
        Polymer & Bulk $\tan\delta$ & Foam & $f_p$ & $\tan\delta_{\rm eff}$ \\ \hline
        Polystyrene &  $6.9\times10^{-3}$ \cite{lt_poly}
        & Styrofoam & $0.023$ & $1.7\times10^{-4}$ \\
        \noalign{\color{gray}\hrule height 0.2pt}
        \multirow{3}{*}{HDPE} & \multirow{3}{*}{$2.5\times10^{-4}$ \cite{James_1996, Llewellyn_1980}} & HD110 & 0.115 & $2.8\times10^{-5}$ \\
         &  & HD80 & 0.083 & $2.1\times10^{-5}$ \\
         &  & HD30 & 0.030 & $8.2\times10^{-6}$ \\
        \noalign{\color{gray}\hrule height 0.2pt}
        \multirow{2}{*}{LDPE} & \multirow{2}{*}{$2.7\times10^{-4}$ \cite{James_1996, Llewellyn_1980}} & LD24 & 0.025 & $7.3\times10^{-6}$ \\
         &  & LD15 & $0.015$ & $4.5\times10^{-6}$ \\ 
        \hline\hline
    \end{tabular}
    \caption{Loss tangent estimates in decreasing order derived independently from our model.}
    \label{tab:eff_lt}
\end{table}

This approach accounts for the redistribution of electric field energy within the composite, which preferentially resides in the low-permittivity nitrogen phase and suppresses dissipation within the polymer matrix.  In the absence of field redistribution, the naive volume-weighted loss tangent estimate is $\tan\delta_{\rm eff} \approx f_p \tan\delta_p + f_m \tan\delta_m \approx~f_p~\tan\delta_p$. Since the polymer loss tangent is many orders of magnitude above the loss tangent of air, $f_m\tan\delta_m \ll f_p\tan\delta_p$ holds yielding $\tan\delta_{\rm eff} \approx f_p \tan\delta_p$. Comparing this estimate with Eq.~\ref{eq:eff_lt}, the MG field redistribution correction is ${\sim}1.1$ for these polymer foams, indicating that the absorption is governed primarily by the bulk material properties. 


\subsection{Connection to Scattering Theory}
\label{sec:cst}
To compare our model to scattering theory, we must estimate the necessary material properties. We approximate the number density of a sample as
\begin{equation}
    N \approx \frac{f_m}{V} = \frac{6f_m}{\pi d^3} \label{eq:num_density}
\end{equation}
for average cell volume $V$ and average cell diameter $d$ using the fractional matrix void volume $f_m$ found in \ref{sec:bp}.

We then combine the geometrical cross section $G$ and number density to define the geometric prefactor relating the scattering efficiency to the attenuation,

\begin{equation}
    NG \approx \frac{3f_m}{2d} \label{eq:NG}.
\end{equation}

Finally, we estimate the mean wall thickness $\ell$ assuming thin spherical cells as
\begin{equation}
    \ell = \frac{d}{2}\left[1-f_m^{1/3}\right].
\end{equation}

For 400 micron cells with a fractional polymer volume $f_p = 0.03$ (reasonable estimates for both Styrofoam and HD30), we find $N \approx 3\times10^{4} \text{ cells/cm}^3$, $NG \approx$ 40 cm$^{-1}$, and $\ell \approx 2$ microns. 

We then invert the model scattering parameters to find the effective length scales governing each scattering pathway. Using Eqs.~(\ref{eq:att}) and (\ref{eq:NG}), we connect Eqs.~(\ref{eq:alpha_M}) and (\ref{eq:qrgd_shell}) for Mie along with Eqs.~(\ref{eq:alpha_R}) and Eqs.~(\ref{eq:qray_shell}) for Rayleigh finding
\begin{align}
    d &= \frac{c(m-1)K_{\rm Mie}}{n_m\pi \sqrt{6} K_{\rm Ray}^2}
         \left|\frac{(n_p^2+2)(1+2n_p^2)-2(n_p^2-1)^2}{(1+2n_p^2)(n_p^2-1)}\right|,\\
    \ell &= \frac{c}{2n_m \pi(m-1)K_{\rm Mie}}\sqrt{\frac{Cd}{6 f_m}}.
\end{align}

Evaluating with $f_m = 0.97$, $n_m \approx 1$, and $n_p = 1.54$, and the fitted parameters for Zotefoam HD30 Batch 3 (Table~\ref{tab:params}), we find $d \approx 200~$\textmu m, $\ell \approx 0.8$~\textmu m. While the prefactor $C$ is responsible for encoding shape and size irregularities, the thin-shell scattering suppression effect, number density, and geometric cross section, the distribution of cell shapes and sizes (polydisperion) is not considered within the utilized theoretical scattering framework (see appendices~\hyperref[appendix:A]{A} and \hyperref[appendix:B]{B} for a full description). Therefore, while the ratio between length scales is essentially correct, these recovered length scales have a factor-of-two disagreement with microscopic values which we attribute to polydisperion. These inversions hold for other batches as well, but break down for denser Zotefoams (HD80 and HD110) which are no longer in the thin shell regime and yield unphysical estimates.

\section{Discussion}  
\label{sec:discussion}

\paragraph{Millimeter} We show the modeled transmittance extrapolated to a 5~cm thick filter for all measured samples in Fig.~\ref{fig:5cm_spectra}.  The most notable performance difference is between Styrofoam and Zotefoams, with further optimization possible via selecting a low density Zotefoam and identifying a high transmittance batch. Following this procedure, the second LD24 batch was selected for 220/280 GHz observations in an SO SAT.

Fig.~\ref{fig:components} separates the transmittance losses by model component, summarizing the in-band performance via a width-normalized scattering. Dielectric absorption is derived from the model parameters shown in Table~\ref{tab:params}. While Styrofoam does not deform under vacuum, the larger in-band scattering and absorption of styrofoam make Zotefoams a better filter material from an optical perspective.

\begin{figure}[h]
    \centering
    \includegraphics[width=\linewidth]{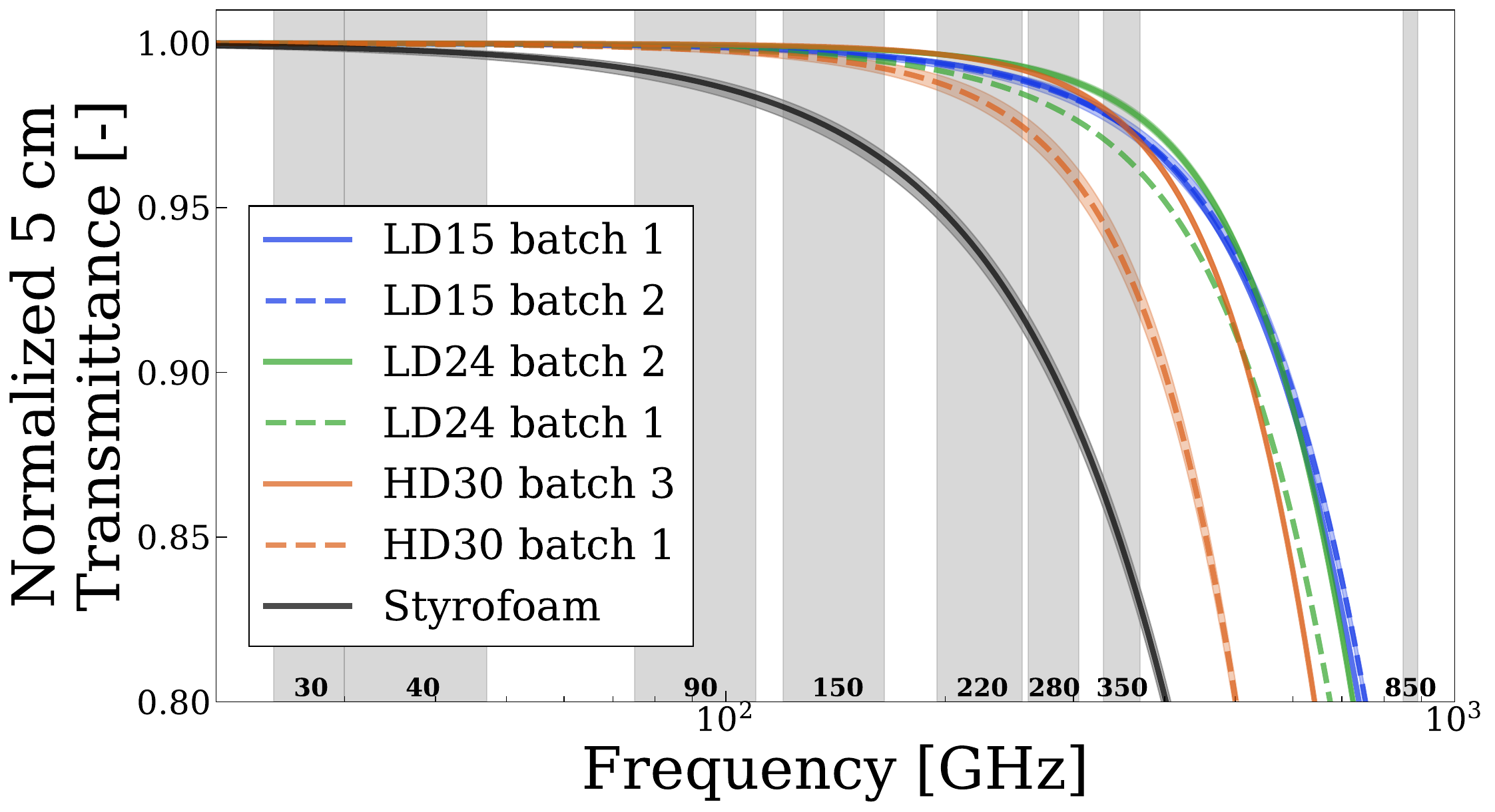}
    \caption{Transmittance models for example 5~cm thick filters. Solid (dashed) lines represent the highest (lowest) transmittance batch of a given foam; shaded regions are the 2$\sigma$ bounds.
    Zotefoam transmittance generally increases as density decreases.
    Styrofoam performs worse than the worst Zotefoam HD30 across all viable bands. Based on this result, a 220/280 GHz Styrofoam filter was replaced with an LD24 batch 2 filter (solid green), increasing transmittance.}
    \label{fig:5cm_spectra}
    \vspace{-1em}
\end{figure}

Following the multi-layer filter approximation, higher density Zotefoams (HD80 and HD110) are not expected to increase IR attenuation while the in-band losses degrade observations. LD24 and LD15 are strongly preferred over a typical HD30 or Styrofoam batch above 200 GHz. These Zotefoam materials remain ideal filters in the single-scattering regime below $\sim$500~GHz; for 850~GHz observations metal mesh filters \cite{Ade2006} are superior.

\begin{figure*}[p]
    \center
    \includegraphics[width=\linewidth]{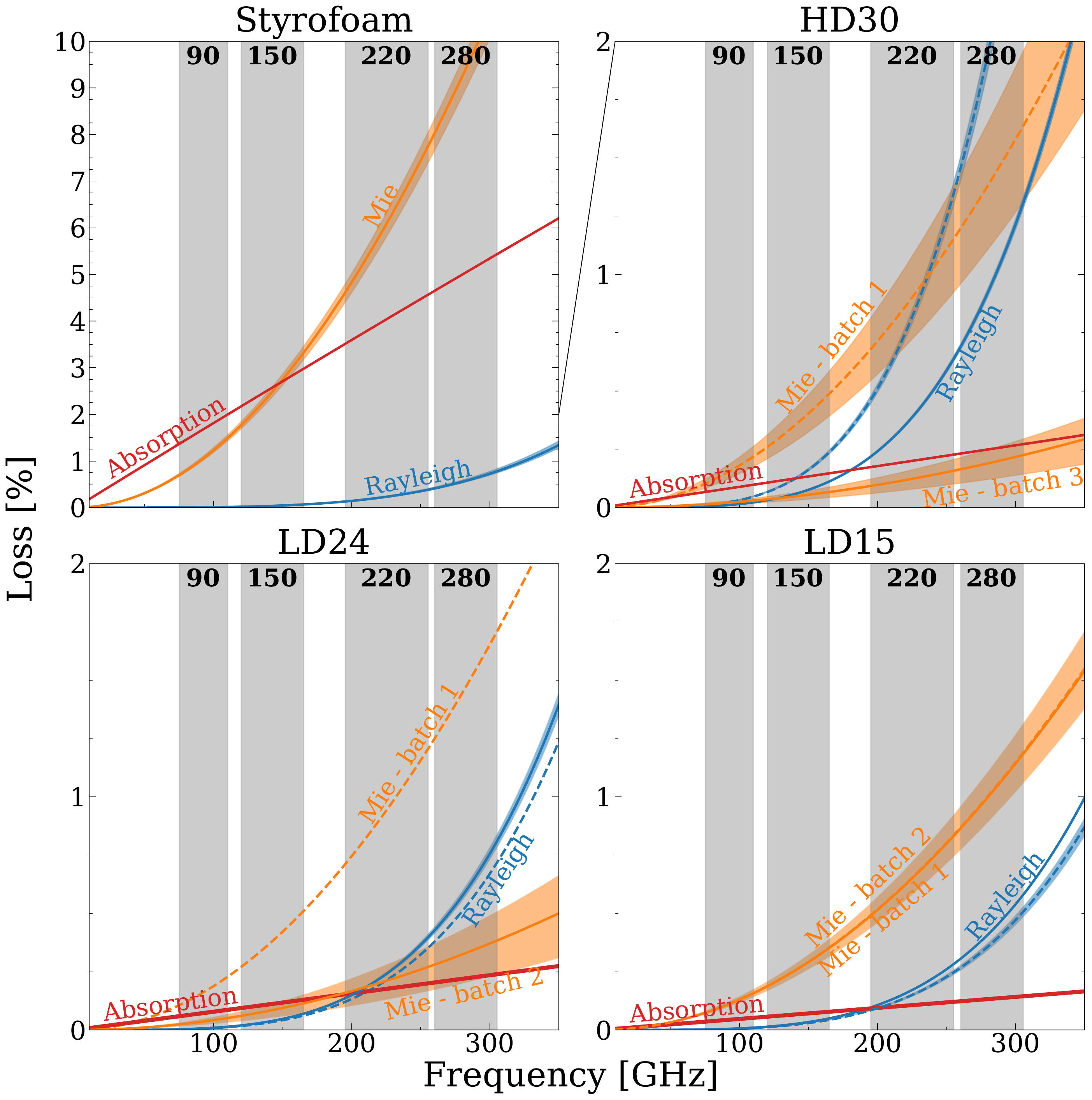}
    \caption{The intrinsic absorption and extrinsic scattering losses of polymer foams filters using a typical 5~cm thickness. Best (worst) batches use solid (dashed) lines. Shaded regions are $1\sigma$ bounds. Absorption is estimated from material properties and uncertainties are therefore not shown. Grey regions are typical atmospheric band windows. Secondary scattering is negligible below 500 GHz.\\
    \textit{Upper Left:} The Styrofoam components have the largest scatter, and the largest estimated absorption via Maxwell-Garnett effective-medium theory. Both degrade mapping speed through attenuating the mm-wave signal and increasing optical loading.\\
    \textit{Upper Right:} All three tested HD30 batches are superior to even the best performing Styrofoam sample. The third batch exhibits the least scatter in both pathways, indicating a smaller average cell size and cell wall thickness.
    We likewise estimate the HD30 loss tangent yielding a ${\lesssim}0.2\%$ decrease in overall transmittance and minimal optical loading.\\
    \textit{Bottom:} LD24 (left) and LD15 (right) are low density Zotefoams. While the performance is generally superior to HD30, batch-to-batch variability persists. Absorption is estimated in an identical procedure.}
    \label{fig:components}
\end{figure*}


Scattering and absorption in return yield increased noise through optical loading. We propagate the re-radiated emission due to absorption to find the resulting optical loading per transition edge sensor (TES, \cite{TES}) detector,
\begin{equation}
   P_{\rm ems} = k_B T_{\rm RJ} \Delta\nu =k_B\langle\alpha_\delta \rangle_\nu T_{\rm eff}\Delta\nu \label{eq:ol_ems}
\end{equation}
for bandwidth $\Delta\nu$ and Rayleigh-Jeans brightness temperature $T_{\rm RJ}$, or effective filter temperature $T_{\rm eff}$. We define $T_{\rm eff}$ as the width-weighted average temperature across the filter by assuming negligible attenuation and reflections. We report optical loadings in Table~\ref{tab:optical_loading} at 273~K as a conservative upper bound.

\begin{table*}[h!]
\centering
\renewcommand{\arraystretch}{1.4}
\begin{tabular}{l ccc ccc}
\hline\hline
& \multicolumn{3}{c}{\textbf{90 GHz}} & \multicolumn{3}{c}{\textbf{220 GHz}} \\[-0.5em]
& \multicolumn{3}{c}{\textbf{150 GHz}} & \multicolumn{3}{c}{\textbf{280 GHz}} \\
\cmidrule(lr){2-4}\cmidrule(lr){5-7}
Sample
& $P_{\rm scatter}$ [pW] & $P_{\rm ems}$ [pW]$^*$ & $P_{\rm tot}$ [pW] & $P_{\rm scatter}$ [pW] & $P_{\rm ems}$ [pW]$^*$ & $P_{\rm tot}$  [pW] \\
\hline
\multirow{2}{*}{Styrofoam} & $0.50^{+0.02}_{-0.02}$ & $2.21$ & $3^{+2}_{-2}$ & $4.6^{+0.2}_{-0.1}$ & $6.84$ & $11^{+7}_{-7}$ \\
 & $1.61^{+0.07}_{-0.05}$ & $4.37$ & $6^{+4}_{-4}$ & $11.2^{+0.4}_{-0.3}$ & $12.34$ & $20^{+10}_{-10}$ \\[6pt]
 HD30 \\[-0.5em]
\multirow{2}{*}{\qquad batch 1} & $0.09^{+0.01}_{-0.01}$ & $0.11$ & $0.2^{+0.1}_{-0.1}$ & $1.2^{+0.1}_{-0.1}$ & $0.34$ & $1.5^{+0.4}_{-0.4}$ \\
 & $0.32^{+0.05}_{-0.05}$ & $0.22$ & $0.5^{+0.2}_{-0.2}$ & $3.4^{+0.3}_{-0.3}$ & $0.62$ & $4.0^{+0.7}_{-0.7}$ \\[6pt]
\multirow{2}{*}{\qquad batch 2} & $0.133^{+0.008}_{-0.008}$ & $0.11$ & $0.2^{+0.1}_{-0.1}$ & $1.42^{+0.08}_{-0.07}$ & $0.34$ & $1.8^{+0.3}_{-0.3}$ \\
 & $0.45^{+0.03}_{-0.03}$ & $0.22$ & $0.7^{+0.2}_{-0.2}$ & $3.7^{+0.2}_{-0.2}$ & $0.62$ & $4.4^{+0.6}_{-0.6}$ \\[6pt]
\multirow{2}{*}{\qquad batch 3$^\dagger$} & $0.015^{+0.004}_{-0.003}$ & $0.11$ & $0.1^{+0.1}_{-0.1}$ & $0.33^{+0.03}_{-0.03}$ & $0.34$ & $0.7^{+0.3}_{-0.3}$ \\
 & $0.071^{+0.012}_{-0.010}$ & $0.22$ & $0.3^{+0.2}_{-0.2}$ & $1.04^{+0.08}_{-0.07}$ & $0.62$ & $1.7^{+0.6}_{-0.6}$ \\[6pt]
 LD24 \\[-0.5em]
\multirow{2}{*}{\qquad batch 1} & $0.0805$ & $0.10$ & $0.18^{+0.10}_{-0.10}$ & $0.835$ & $0.30$ & $1.1^{+0.3}_{-0.3}$ \\
 & $0.271$ & $0.19$ & $0.5^{+0.2}_{-0.2}$ & $2.17$ & $0.54$ & $2.7^{+0.5}_{-0.5}$ \\[6pt]
\multirow{2}{*}{\qquad batch 2$^\dagger$} & $0.021^{+0.007}_{-0.006}$ & $0.10$ & $0.12^{+0.10}_{-0.10}$ & $0.30^{+0.06}_{-0.05}$ & $0.30$ & $0.6^{+0.3}_{-0.3}$ \\
 & $0.08^{+0.02}_{-0.02}$ & $0.19$ & $0.3^{+0.2}_{-0.2}$ & $0.9^{+0.2}_{-0.1}$ & $0.54$ & $1.4^{+0.6}_{-0.6}$ \\[6pt]
 LD15 \\[-0.5em]
\multirow{2}{*}{\qquad batch 1} & $0.0558$ & $0.06$ & $0.11^{+0.06}_{-0.06}$ & $0.591$ & $0.18$ & $0.8^{+0.2}_{-0.2}$ \\
 & $0.190$ & $0.11$ & $0.3^{+0.1}_{-0.1}$ & $1.55$ & $0.33$ & $1.9^{+0.3}_{-0.3}$ \\[6pt]
\multirow{2}{*}{\qquad batch 2$^\dagger$} & $0.056^{+0.006}_{-0.006}$ & $0.06$ & $0.11^{+0.06}_{-0.06}$ & $0.58^{+0.05}_{-0.05}$ & $0.18$ & $0.8^{+0.2}_{-0.2}$ \\
 & $0.19^{+0.02}_{-0.02}$ & $0.11$ & $0.3^{+0.1}_{-0.1}$ & $1.5^{+0.1}_{-0.1}$ & $0.33$ & $1.8^{+0.4}_{-0.4}$ \\[6pt]
\multicolumn{7}{l}{\footnotesize* Loss tangent estimate used to bound loading due to a non-detection, propagated error taken to be order unity} \\[-0.5em]
\multicolumn{7}{l}{\footnotesize$\dagger$ Lowest optical loading batch of a given formulation} \\
\hline\hline
\end{tabular}
\caption{Per-TES optical loading estimates for 90 - 280 GHz bands (indicated by the top two rows).
The scattering optical loading considers both the Rayleigh and Mie scattering of the warm forebaffle (taken to be from $\theta_{\rm min} = 40^\circ$ and $\theta_{\rm max} = 90^\circ$) using Eq.~\ref{eq:ol_scatter}. As our fits do not detect absorption, and scattering yields less optical loading than absorption for a given transmittance loss, the scattering optical loadings are a lower bound. We combine the scattered power with the emitted power using the estimated dielectric loss and an upper bound temperature of $T = 273$~K to provide an upper bound on the optical loading of the materials.
Values are band-averaged using square bands of 35, 45, 45, and 65 GHz wide, for 90, 150, 220, 280 GHz bands, respectively.}
\label{tab:optical_loading}
\end{table*}


We estimate the optical loading from the thermal emission of the warm baffle scattered onto the focal plane by these polymer foams. We calculate this per band and per TES as
\begin{align}
    P_{\rm scatter} = k_B T_{\rm amb} \Delta\nu \int_{\theta_{\rm min}}^{\theta_{\rm max}} \big[ &\langle(1-T_{\rm Mie})K_{\rm Mie}(\theta) \rangle_\nu \label{eq:ol_scatter}\\
     + & \langle(1-T_{\rm Ray}) K_{\rm Ray}(\theta)\rangle_\nu \big]\sin\theta d\theta,  \nonumber
\end{align}

where $T_{\rm amb}$ is the ambient temperature, $T_{\rm Mie}$ and $T_{\rm Ray}$ are the Mie and Rayleigh transmittances, $K_i$ are the scattering kernels (found in Appendix~\hyperref[appendix:C]{C}), and the bounds on $\theta$ correspond to the angular extent of the warm baffle as seen by the filter. As this integral is less than unity by forebaffle geometry and scattering kernel normalization, a given fractional absorption always produces larger optical loading than the same fractional scattering loss when $T_{\rm amb}/T_{\rm eff}$ is of order unity.

The largest batch-to-batch difference is $0.8\%$ and $1.8\%$ in Rayleigh and Mie scattering loss, respectively, of HD30. Both are between the first and third batch. We report the batch-to-batch variability without our systematic errors as the unknown statistical uncertainty from our small population of batches dominates as shown in Fig.~\ref{fig:components}.
The LDPE materials (LD24 and LD15) offer lower scatter and absorption. Though batch-to-batch variability is a confounding factor, LD24 and LD15 materials can offer significant improvements over HD30. Given the percent-level scattering variability between batches across all foam formulations, identifying an optically pristine batch is key for optimization and feasible given the cost-effective nature of these foams.

Based on our scattering measurements and absorption estimates, we use \textit{jbolo}\footnote{\url{https://github.com/JohnRuhl/jbolo}, derived from BoloCalc \cite{bolocalc}} to predict {$\sim$}16\%, {$\sim$}19\%, {$\sim$}35\%, {$\sim$}38\% increased mapping speed from replacing the SO SAT 4.5~cm thick Styrofoam filters with a 5~cm thick LD24 filter for 90, 150, 220, and 280~GHz observations, respectively.
While Styrofoam has been shown to be sufficient for 90/150 observations \cite{Day-Weiss_2026, Harrington_2026}, future replacements in 90/150 GHz SO SATs may follow if the predicted mapping speed gains are realized at 220/280 GHz.

\paragraph{Infrared} 

The net radiative IR flux $q$ (W/m$^2$) is approximated as radiative exchange between $N$ intermediate layers assuming sufficient absorption for thermalization due to vibrational polymer modes, a net radiative equilibrium, local thermal equilibrium at each layer, and homogeneous emission of order unity,
\begin{align}
q_{\mathrm{load}} = \frac{\sigma}{N + 1}
\left[\left(T_{\mathrm{high}}^4 - T_{\mathrm{low}}^4\right)
-\varepsilon\sum_{i=1}^N\left(T_{\textrm{top},i}^4 - T_{\textrm{bot},i}^4\right) \right]
\label{eq:MLE}
\end{align}
where $\sigma$ is the Stefan-Boltzmann constant, $T_{\mathrm{high}}$ and $T_{\mathrm{low}}$ are the equilibrium temperatures above and below the filter stack, respectively, $\varepsilon$ is the effective emissivity, and $T_{\textrm{top},i}$ and $T_{\textrm{bot},i}$ are the top and bottom surface temperature of the $i$-th layer \cite{Choi_2013}. This approximation neglects conduction between layers, geometric and transverse radiative effects, as well as forward coupling to cell-wall structures in the far-infrared \cite{EDW_notes}. 


The number of layers is set by system-level optimization which considers the IR attenuation at colder stages as well. The sheet thicknesses, chosen to be IR opaque, determines the total filter thickness and thus the in-band optical loading. Filters are typically constructed from the thinnest available sheet stock, with thicker sheets introduced as needed for mechanical robustness against buckling or sagging \cite{Day-Weiss_2026, Sobrin_2022}. Overly thin sheets can be highly transparent in the far-IR \cite{helson_2025}, resulting in partial transmission and thermal loading above the radiative exchange approximation.

If thermal conduction is negligible, the temperature $T$ of layer $i$ is
\begin{equation}
    T_i = \left[ T_\mathrm{high}^4 - \frac{i}{N+1}
    \left( T_\mathrm{high}^4 - T_\mathrm{low}^4 \right)
    \right]^{1/4}. \label{eq:MLE_i}
\end{equation}

Thermal conductance across each layer decreases thermal loading by allowing a temperature gradient ($T_{\textrm{top},i} > T_{\textrm{bot},i}$) which results in decreased emission towards the focal plane. Using Eq.~\ref{eq:MLE} with $T_{\textrm{high}}~=~273$~K, $T_{\textrm{low}}~=~44$~K, an area of $0.37~\textrm{m}^2$, and solving the radiative transfer equilibrium to estimate the top and bottom surface temperature of each layer predicts 4~W of thermal loading for a 4.6~cm thick, 24 layer Styrofoam filter and 5~W without conduction (where $T_{\textrm{top},i} = T_{\textrm{bot},i}$). Similarly, using Eq.~\ref{eq:MLE}, predicts a comparable 6~W of thermal loading with conduction and 7~W without conduction for the 5~cm, 16 layer LD24 filter. We solve the layers temperatures using thermal conductivities of 0.028 for Styrofoam and 0.036 for LD24 \cite{Choi_2013, LD24}

However, thermal conductance between layers decreases the thermal contrast between layers ($T_{\textrm{top},i+1} \rightarrow T_{\textrm{bot},i}$) and increases thermal loading, resulting in up to 20~W for both filters in limit of a one-layer conductive slab. For deployment without spacers reducing thermal contact between layers, the low thermal conductance of these materials is therefore ideal. Day-Weiss \textit{et al.} \cite{Day-Weiss_2026} confirms excess thermal loading, measuring ${<}12$~W for a 4.6~cm thick, 24 layer Styrofoam filter operating within an SO SAT which does not use spacers. In addition to the known thermal contact between layers, this excess is attributed to rejected power being reflected back into the instrument by aluminum filter holder. 
HD30 has been widely used for IR blocking within the CMB community; the deployed LD24 filter is expected to offer similar IR blocking to HD30 based on the similar cell sizes and shared polyethylene absorption features.

\paragraph{Mechanical Properties}
While Styrofoam can be cryogenically cycled without replacement, Zotefoam HD30 is known to require glue or fasteners to prevent deformation under cryogenic cycling and does not have the same lifespan. Lower density Zotefoams have increased deformability that may further shorten their lifespans. Three rounds of LD24 vacuum testing were performed and cryogenic LD24 testing is underway within an SO SAT. While future deployment of LD15 is appealing, it has not been mechanically tested, a risk that must be retired before adoption.

\section{Conclusion}
\label{sec:conclusion}

We presented broadband transmittance spectroscopy of Styrofoam and five Zotefoam formulations from 150 GHz to 2 THz, fitting each spectrum with a radiative transfer model incorporating Rayleigh scattering, Mie scattering, dielectric absorption, and a coupled secondary scattering correction. Our radiative transfer framework distinguishes between absorptive and scattering losses. Each model component is constrained at the ${\sim}0.1\%$ transmittance level, enabling quantitative decomposition of loss mechanisms across materials and bands. 

Of the possible secondary scattering channels, only the Rayleigh–Mie coupling is detected, consistent with the cell-matrix structure. We use Maxwell–Garnett effective-medium theory to provide an upper bound for the undetected Zotefoam loss tangent, constraining absorption and thermal emission. 
Connections to the microphysical structures are limited by shape irregularities of the cells, which remains poorly estimated, but derives length scales that only differ by a factor of a few from microscopic estimates. Extending this framework to in-situ transmittance measurements to constrain thermal contraction of the cells, and characterizing how material properties such as cell shape and size statistics contribute to batch-to-batch variability, would close the remaining gap between laboratory characterization and predictive end-to-end optical modeling.

While Styrofoam has superior IR properties and a longer lifespan, Styrofoam exhibits high scattering loss and more than an order of magnitude larger loss tangent. This yields optical loading comparable to the atmospheric emission for 220/280 GHz observation 
degrading sensitivity. By contrast, even most scattering Zotefoam HD30 batch has superior mm-wave properties that drive sensitivity to reach the ambitious science goals of modern CMB experiments. For HD30 we find ${<}3\%$ scattering and ${<}0.3\%$ absorption, with batch-to-batch variance of up to {${\sim}2\%$} that can be characterized to select the best batch for further optimization. Lower-density formulations (LD24 and LD15) can further reduce in-band scattering to ${<}1\%$ while maintaining negligible absorption. These findings motivated the deployment of a filter constructed from the best-performing LD24 batch replacing Styrofoam in the Simons Observatory 220/280 Small Aperture Telescope. As we predict tens-of-percent increased sensitivity from this replacement, future low density Zotefoam filters may be deployed in mm-wave cameras if the predicted sensitivity gains are realized in the field.


\appendix
\section*{Appendix A: Scattering Background}
\label{appendix:A}

\paragraph{Attenuation} For a homogeneous material, the spectral transmittance is described by the Beer–Lambert law,
\begin{equation}
    T(\lambda) = \exp[-w\alpha(\lambda)], \label{eq:bl}
\end{equation}
where $w$ denotes the sample thickness and $\alpha(\lambda)$ the wavelength-dependent attenuation (also known as extinction $\mu(\lambda)$). 

The attenuation for a population of spherical scatterers with number density $N$ can be defined in terms of the single-particle cross section $\sigma(\lambda)$, geometric cross section $G$, and scattering efficiency $Q(\lambda)$, or mean free path $\lambda_{\rm mfp}$
\begin{equation}
    \alpha(\lambda) \equiv N\sigma(\lambda) \equiv N G Q(\lambda) \equiv \lambda_{\rm mfp}^{-1}(\lambda).
    \label{eq:att}
\end{equation}

\noindent When multiple loss mechanisms contribute, the transmittance is the product of the individual terms, which is equivalent to the sum of attenuations,
\begin{align}
    T(\lambda) = \prod_i T_i(\lambda), \qquad \alpha(\lambda) = \sum_i \alpha_i(\lambda).
\end{align}
This multiplicative property of the transmittance processes underlies the treatment of Rayleigh, Mie, and dielectric losses in the main text.

\paragraph{Scattering efficiencies} The scattering efficiency is defined as
\begin{equation}
    Q = \frac{2}{x^2}\sum_{n=1}^{\infty}(2n+1)\left(|a_n|^2 + |b_n|^2\right)
    \label{eq:q}
\end{equation}
for size parameters $x \equiv kd/2 = n_m \pi d/\lambda$, wavenumber $k$, cell diameter $d$, and electric and magnetic Mie coefficients $a_n$ and $b_n$ using the Riccati-Bessel functions $\psi_n$ and $\xi_n$,
\begin{align}
a_n &= \frac{ m \,\psi_n(m x)\, \psi_n'(x) - \psi_n(x)\, \psi_n'(m x) }
           { m \,\psi_n(m x)\, \xi_n'(x) - \xi_n(x)\, \psi_n'(m x) }, \label{eq:a_n} \\
b_n &= \frac{ \psi_n(m x)\, \psi_n'(x) - m \,\psi_n(x)\, \psi_n'(m x) }
           { \psi_n(m x)\, \xi_n'(x) - m \,\xi_n(x)\, \psi_n'(m x) } \label{eq:b_n}
\end{align}

In the Rayleigh limit $x \ll 1$, $a_n \propto x^{2n+1}$ and $b_n \propto x^{2n+3}$, higher order multiples are negligible and only the electric dipole remains resulting in the well-known $f^4$ dependence,
\begin{equation}
        Q_{\textrm{Ray}}(x) = \frac{8}{3}x^4\left|\frac{m^2-1}{m^2+2}\right|^2. \label{eq:qray}
\end{equation}

For larger particles where the accumulated phase error across the wavefront is
non-negligible, the anomalous diffraction approximation
\cite{vdH} treats each ray traversing the sphere as acquiring a phase
shift $\rho = 2x(m-1)$ and applies the optical theorem to obtain $Q_{\rm Mie}(\rho)$.
\begin{equation}
    Q_{\textrm{Mie}}(\rho) = 2 - \frac{4}{\rho}\sin(\rho)+\frac{4}{\rho^2}(1-\cos(\rho)). \label{eq:qmie}
\end{equation}
In the limit $\rho \ll 1$, a Taylor expansion recovers the Rayleigh-Gans-Debye approximation $Q_{\rm RGD}$,
\begin{equation}
    Q_{\textrm{RGD}}(\rho) = \frac{1}{2}\rho^2, \label{eq:qrgd}
\end{equation}

\noindent The dominance of Mie scattering at small frequencies ($\rho \ll 1$) then follows directly, $Q_{\rm Ray}/Q_{\rm RGD} \propto f^2$. In the opposite regime regime ($\lambda_{\rm mfp} \ll  w$), the vanishing mean free path leads to many scattering events of both kinds producing an effectively randomized exit angle and an optically white appearance (e.g., clouds). This phenomenon, known as the diffuse scattering limit, is relevant for IR blocking in these foams where incident light becomes depolarized and isotropic while the increased optical path length promotes absorption within each layer.

\paragraph{Scattering Kernels}
The angular distribution of Rayleigh and Mie scattering can be described using the normalized scattering kernel $\Phi(\theta,\phi;f)$, commonly referred to as the phase function. For the IR-blocking foams considered here, scattering occurs predominantly from spherical cells. This azimuthal symmetry reduces the kernel to a function of polar angle only,
\begin{equation}
\Phi_i(\theta,\phi;\lambda) = \frac{1}{2\pi}K_i(\theta;\lambda), \qquad \int_0^\pi K_i(\theta;\lambda)\,\sin\theta\,d\theta = 1. \label{eq:Phase_func}
\end{equation}

For an isotropic material that does not induce cross polarization, the kernel can be written in terms of the perpendicular scattering amplitude $S_1(\theta)$ and parallel scattering amplitude $S_2(\theta)$,
\begin{equation}
    K(\theta) = \abs{S_1(\theta)}^2 + \abs{S_2(\theta)}^2
\end{equation}

\noindent where the amplitudes are defined via the infinite series

\begin{align}
S_1(\theta) &= \sum_{n=1}^{\infty} \frac{2n+1}{n(n+1)} \left[ a_n \, \pi_n(\cos\theta) + b_n \, \tau_n(\cos\theta) \right], \\
S_2(\theta) &= \sum_{n=1}^{\infty} \frac{2n+1}{n(n+1)} \left[ a_n \, \tau_n(\cos\theta) + b_n \, \pi_n(\cos\theta) \right]
\end{align}
\noindent with angular functions $\pi_n$ and $\tau_n$ using the associated Legendre polynomial $P_n^1$,

\begin{align}
\pi_n(\cos\theta) = \frac{P_n^1(\cos\theta)}{\sin\theta}, \qquad
\tau_n(\cos\theta) = \frac{d}{d\theta}P_n^1(\cos\theta).
\end{align}

In the Rayleigh scattering limit ($n=1$, $x\ll 1$), only the electric dipole is considered. Therefore, $\pi_1(\cos\theta) = 1$, $\tau_1(\cos\theta) = \cos\theta$, and the only non-vanishing Mie coefficient $a_1$ is removed by normalization yielding the well-known result
\begin{equation}
    S_1 = 1, \qquad S_2 = \cos\theta, \qquad K(\theta) = 1 + \cos^2\theta.
\end{equation}

\section*{Appendix B: Foam Scattering Efficiencies}
\label{appendix:B}

\noindent For expanded closed-cell foams, scattering occurs from thin polymer shells of width $\ell$ surrounding matrix voids of diameter $d$ with $\ell \ll d$. Compared to spherical scatters of the same diameter, the scattering efficiencies in thin shells are strongly suppressed, arising from the same thin-layer expansion of the electromagnetic boundary conditions that govern anti-reflection coating designs \cite{bohren_huffman}. The Mie coefficients for these foams can be derived from \cite{Adan_Kerker} by allowing the inner sphere and surrounding media to have the same refractive index and expanding to first order since scattering vanishes with $\ell\to0$ without dielectric contrast between the interior and exterior void media. Lange \& Aragón \cite{Lange_Aragon} derive the Mie coefficients for thin polymer shells where $\ell \ll d$ in this manner,
\begin{align}
    a_n &= -i k \ell \left[
        \frac{n_p(n_p+1)(m^2 - 1)}{m^2 x^2} \psi_n^2(x)
        + (m^2 - 1)\psi_n^{\prime\,2}(x)
    \right], \label{eq:an_shell} \\
    b_n &= -i k \ell \,(m^2 - 1)\,\psi_n^2(x). \label{eq:bn_shell}
\end{align}

In the Rayleigh limit $x \ll 1$, only the electric dipole $a_1$ contributes at leading order. For nitrogen and for true voids, we set $m = n_p$. Substituting the small-argument expansions $\psi_1(x) \approx x^2/3$ and $\psi_1'(x) \approx 2x/3$ into Eqs.~(\ref{eq:an_shell})--(\ref{eq:bn_shell}) and writing $k\ell = (2\ell/d)x$, the sum in Eq.~(\ref{eq:q}) reduces to
\begin{align}
    Q_{\rm Ray,\,shell}(x) &= \frac{8}{3} x^4 \left(\frac{6\ell}{d}\right)^2 \left|\frac{(1 + 2n_p^2)(n_p^2 - 1)}{9n_p^2}\right|^2, \label{eq:qray_shell} \\
    &= \left(\frac{6\ell}{d}\right)^2 \left|\frac{(1 + 2n_p^2)(n_p^2 +2)}{9n_p^2}\right|^2Q_{\rm Ray}.
\end{align}
The same result can be obtained via the polarizability of an electrostatic coated sphere (see \cite{bohren_huffman}, pg. 149).

At normal incidence, phase errors accumulate only within the polymer shell of total thickness $2\ell$, giving 
$\rho_{\rm shell} = 2k\ell(m-1) = (2\ell/d)\rho$. At oblique incidence, rays traverse a path length of $2\ell/\cos\theta$, producing a divergent contribution to the full anomalous diffraction approximation integral which only worsens agreements with microscopic estimates. We therefore retain only the normal-incidence result and substitute into Eq.~(\ref{eq:qrgd}) to obtain,
\begin{equation}
    Q_{\textrm{RGD,\,shell}}(\rho) \simeq \frac{1}{2}\left(\frac{2\ell}{d}\right)^2 \rho^2 = \left(\frac{2\ell}{d}\right)^2Q_{\textrm{RGD}} \label{eq:qrgd_shell}
\end{equation}
\noindent which carries the same $(\ell/d)^2$ as Eq.~(\ref{eq:qrgd_shell}) but reflects the decreased path length rather than decreased polarizable volume.

\section*{Appendix C: Scattering Kernels}
\label{appendix:C}

We compare Mie scattering kernels for a void sphere in an HDPE matrix versus a void sphere, thin HDPE shell, and void exterior in Fig.~\ref{fig:sim_kernels}. The differences illustrates the impact of an expanded foam with adjoined cells. The Rayleigh scattering kernel is only dependent on the electric dipole, and therefore remains $1 + \cos^2\theta$ up to normalization. In Fig.~\ref{fig:mes_kernels}, we show angular scattering measurements of a 2.54 cm thick Zotefoam HD30 sample which reflects the increasing scattering efficiency with frequency and impact of the scattering kernel. 

\begin{figure}[h]
    \centering
    \includegraphics[width=\linewidth]{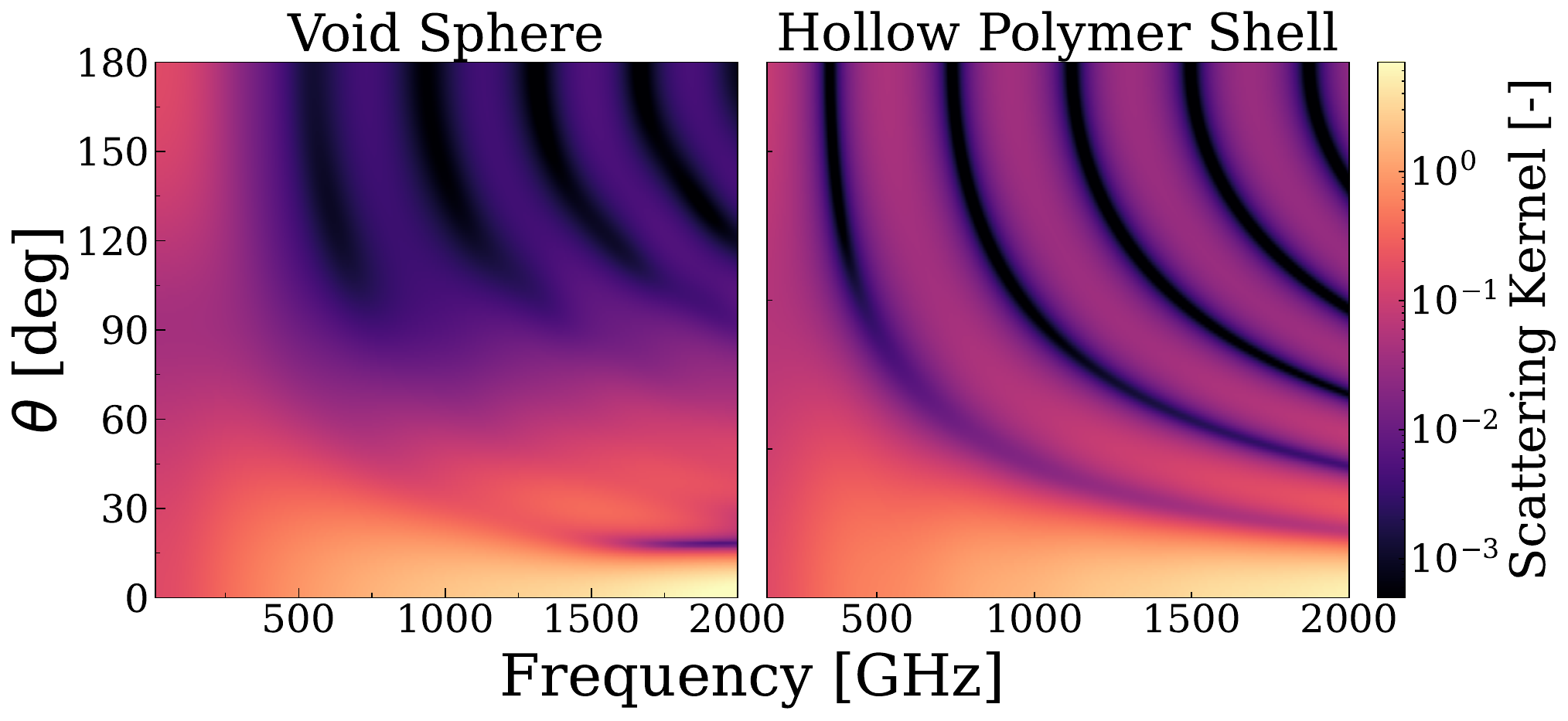}
    \caption{Simulated single Mie scattering kernels for a vacuum sphere suspended in HDPE (left) versus a thin HDPE shell with vacuum interior and exterior (right). The void sphere uses the ordinary Mie coefficients (Eqs.~(\ref{eq:a_n}) and (\ref{eq:b_n})), while the hollow polymer shell uses the Mie coefficients derived for this special case (Eqs.~(\ref{eq:an_shell}), and (\ref{eq:bn_shell})) by \cite{Lange_Aragon}. In both cases, the scattering kernel narrows with increasing frequency. Backscatter and interference effects are suppressed for shells, yielding more prominent Mie resonance patterns. Both cells are 400 microns and the HDPE index is 1.54; the cell walls are 2 microns.}
    \label{fig:sim_kernels}
\end{figure}

\begin{figure}[h!]
    \centering
    \includegraphics[width=\linewidth]{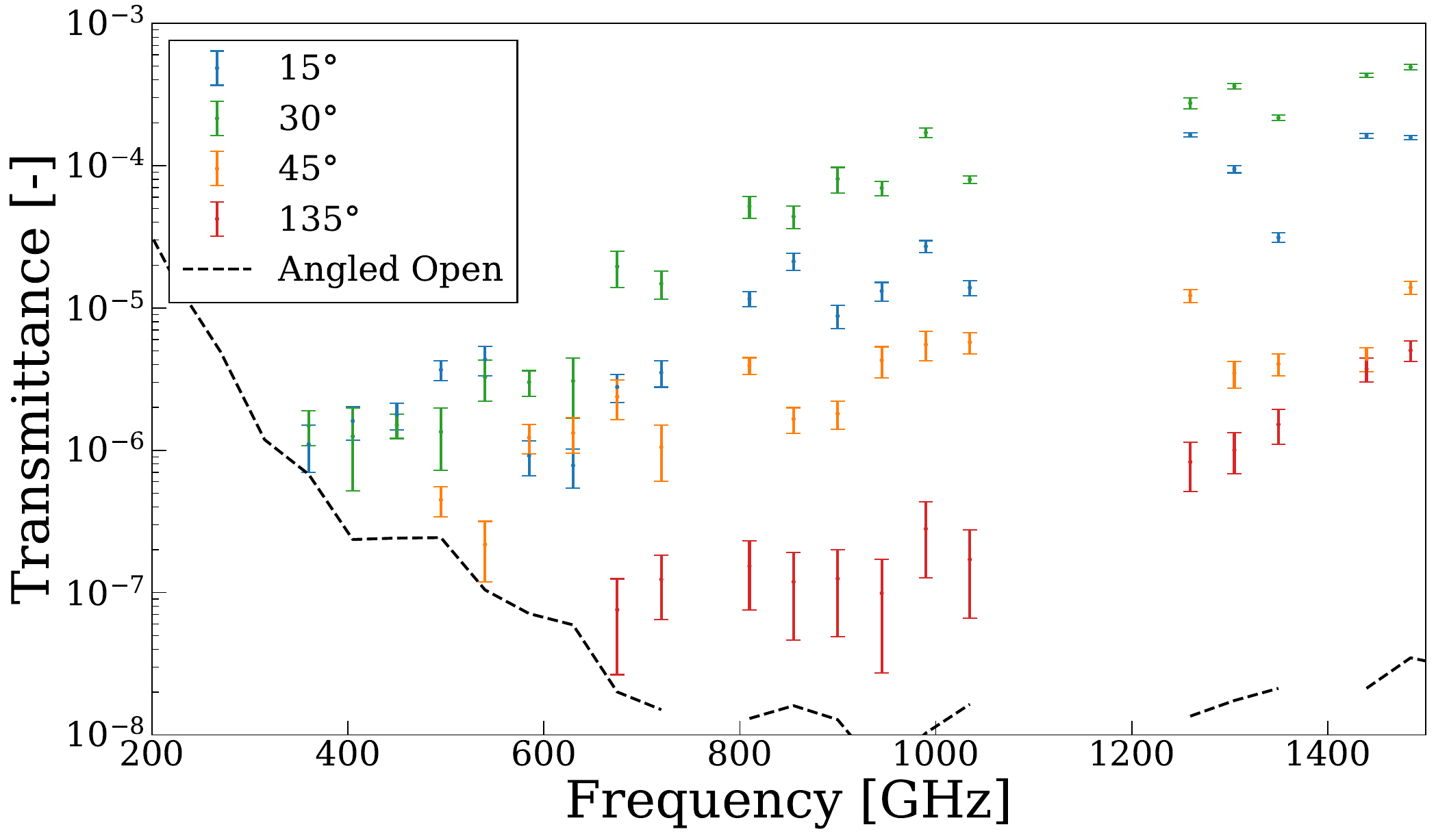}
    \caption{Measured transmittance at various angles arising due to scattering normalized by the in-line transmittance. Wider 50 GHz binning is employed due to lower SNR. Data near the open measurement (no sample; noise floor) at 15 degrees is noise dominated and therefore excluded. Narrower angles generally have higher transmittance as expected, while the 30 degree transmittance exceeding that of 15 degrees could be explained by Mie resonances.}
    \label{fig:mes_kernels}
\end{figure}

\section*{Appendix D: Higher-Order Scattering}
\label{appendix:D}
The correction for multiple scattering may be generalized to arbitrary order. For order $n$, there are $n+1$ independent channels, and the 
$n$th-order contribution is
\begin{equation}
S_{n} = \sum_{m=0}^n \binom{n}{m}\,a_{R^{n-m}M^m}\,\alpha_R^{\,n-m}\,\alpha_M^m C^{1-n},
\end{equation}
where the binomial coefficient $\binom{n}{m}$ accounts for the number of permutations of Rayleigh and Mie scatters in each channel since scattering processes are order independent.

Including absorption from dielectric loss $\alpha_\delta$, the total transmittance 
through order $n$ is
\begin{equation}
T_n = \exp\Big[-w\Big(\alpha_R + \alpha_M + \alpha_\delta - \sum_{k=2}^n S_k\Big)\Big].
\end{equation}

These higher-order corrections may prove useful to accurately model into the highly attenuating, diffuse scattering-dominated regime.

\section*{Appendix E: Waterlines}
\label{appendix:E}
To mitigate contamination from atmospheric absorption and emission features in our free space transmittance spectroscopy measurements, we identify and mask prominent features; all of which are rotational transitions of H$_2$0 as shown in Table~\ref{tab:h2o}.

\begin{table}[h]
    \centering
    \renewcommand{\arraystretch}{1.3}
    \setlength{\tabcolsep}{2pt}
    \begin{tabular}{rcccc}
    \hline\hline
    Frequency [GHz] & Spin Isomer & Transition & Mask [GHz] & Ref. \\
    \hline
    556.936  & ortho & $1_{10} \rightarrow 1_{01}$ & 545 -- 570   & \cite{557} \\
    752.033  & para  & $2_{11} \rightarrow 2_{02}$ & 740 -- 765   & \cite{752} \\
    1113.343 & para  & $1_{11} \rightarrow 0_{00}$ & 1080 -- 1245 & \cite{waterlines} \\
    1162.912 & ortho & $3_{21} \rightarrow 3_{12}$ & $\cdots$ & \cite{waterlines} \\
    1228.789 & para  & $2_{20} \rightarrow 2_{11}$ & $\cdots$ & \cite{waterlines} \\
    1410.618 & ortho & $5_{23} \rightarrow 5_{14}$ & 1390 -- 1430 & \cite{waterlines} \\
    1661.007 & ortho & $3_{12} \rightarrow 3_{03}$ & 1585 -- 1810 & \cite{waterlines} \\
    1669.905 & para  & $2_{21} \rightarrow 2_{12}$ & $\cdots$ & \cite{waterlines} \\
    1884.888 & ortho & $8_{45} \rightarrow 7_{53}$ & 1855 -- 1950 & \cite{waterlines} \\
    \hline\hline
    \end{tabular}
    \caption{The frequency, spin isomer form, transition, mask bounds for our analysis, and reference of the rotational water lines detected in our spectra from 100 to 2,000 GHz.}
    \label{tab:h2o}
\end{table}

\begin{backmatter}
\bmsection{Funding}
This work was supported by the National Science Foundation Graduate Research Fellowship (Grant No. 2140001), the Simons Foundation (Award \#457687, B.K.) and the U.S. National Science Foundation (Award Number: 2153201). 
Work at Argonne was supported under the DOE contract DE-AC02-06CH11357. 
GC acknowledges funding from the European Union (ERC, POLOCALC, 101096035) and from Italy Ministry of Research (MSCA2024\_0000016)

\bmsection{Acknowledgment}
We recognize the help of Austin Stover in capturing the microscope images included in this manuscript and Hrushi Athreya for helpful discussions of loading calculations.


\bmsection{Disclosures}
The authors declare no conflicts of interest.

\bmsection{Data Availability Statement}
The general-use TeraScan and \mbox{TeraFlash} processing code \texttt{TeraToptica} \cite{TeraToptica} along with the data, plots, and analysis presented in this work can be found on Zenodo \cite{irbf_data}.

\bmsection{Supplemental document}

\end{backmatter}

\bibliography{bib}

\end{document}